%% file: main.tex
\pdfoutput=1
\documentclass[journal]{IEEEtran}

\input{preamble}
\hypersetup{hidelinks}%

\begin{document}

\raggedbottom

\input{part_manuscript}

\end{document}

%% file: preamble.tex
\usepackage{amsmath,amssymb,amsfonts}
\usepackage{bm}
\usepackage{graphicx}
\graphicspath{{figures/}}
\usepackage{booktabs}
\usepackage{multirow}
\usepackage{array}
\usepackage{algorithmic}
\usepackage{algorithm}
\usepackage{url}
\usepackage{cite}
\usepackage[colorlinks=true,linkcolor=blue,citecolor=blue,urlcolor=blue]{hyperref}
\usepackage{cleveref}
\usepackage{xcolor}
\usepackage{pifont}%
\usepackage{tikz}
\usetikzlibrary{arrows.meta,positioning,fit,backgrounds,calc}
\usepackage{longtable}

\input{math_commands}

\newcommand{\cmark}{\ding{51}}%
\newcommand{\xmark}{\ding{55}}%
\newcommand{\pmark}{{\small$\circ$}}%

%% file: math_commands.tex
\newcommand{\Czero}{\mathsf{C0}}
\newcommand{\Cone}{\mathsf{C1}}
\newcommand{\Ctwo}{\mathsf{C2}}

\newcommand{\statetuple}{\boldsymbol{s}_t}
\newcommand{\riskt}{r_t}
\newcommand{\deadlinet}{d_t}
\newcommand{\bwt}{b_t}

\newcommand{\conflictt}{c_t}
\newcommand{\staleness}{\sigma_t}
\newcommand{\reversibility}{\rho_t}

\newcommand{\taucommit}{\tau_{\mathrm{commit}}}
\newcommand{\taureject}{\tau_{\mathrm{reject}}}
\newcommand{\taudegraded}{\tau_{\mathrm{degraded}}}
\newcommand{\deltastale}{\delta}
\newcommand{\epstrust}{\varepsilon_{\mathrm{trust}}}

\newcommand{\COMMIT}{\textsc{Commit}}
\newcommand{\GATE}{\textsc{Gate}}
\newcommand{\REJECT}{\textsc{Reject}}
\newcommand{\HUMANGATE}{\textsc{Human-Gate}}

\newcommand{\GRNONE}{\textsc{None}}
\newcommand{\RDIV}{\textsc{Risk-Divergence}}
\newcommand{\LCONFL}{\textsc{Local-Conflict}}
\newcommand{\PUPG}{\textsc{Planner-Upgrade}}
\newcommand{\MRISK}{\textsc{Mid-Risk}}

\newcommand{\pp}{\,\mathrm{pp}}

%% file: part_manuscript.tex
\title{Executor-Side Progressive Risk-Gated Actuation\\for Agentic AI in Wireless Supervisory Control}

\author{Zhenyu~Liu, Yi~Ma, and Rahim~Tafazolli%
\thanks{Z. Liu, Y. Ma, and R. Tafazolli are with the 6GIC, Institute
for Communication Systems, University of Surrey, Guildford, United
Kingdom, GU2 7XH (e-mail: \{zhenyu.liu, y.ma, r.tafazolli\}@surrey.ac.uk).}
}

\maketitle

\begin{abstract}
\input{sections/0_abstract}
\end{abstract}

\begin{IEEEkeywords}
Agentic AI, executor-side actuation, O-RAN, runtime assurance, wireless supervisory control
\end{IEEEkeywords}

\input{sections/1_introduction}
\input{sections/2_related_work}

\input{sections/3_system_model}

\input{sections/4_profile_policy}

\input{sections/5_online_selection}

\input{sections/6_calibration}

\input{sections/6_evaluation}

\input{sections/7_discussion}
\input{sections/8_conclusion}

\bibliographystyle{IEEEtran}
\bibliography{references}

%% file: sections/0_abstract.tex
Agentic artificial intelligence built on large language models shows
promise for automating Open Radio Access Network supervisory control.
However, action-level safeguards alone may be insufficient: evidence
acquisition consumes time and shared resources, and network changes can
invalidate approvals before release. We propose Progressive Risk-Gated
Actuation (PRGA), an executor-side model and contract governing intents
from assessment to release or termination. PRGA combines local checks
with selective retrieval of coordination evidence to reduce acquisition
costs while preserving mandatory evidence requirements. Version and
lease rechecks prevent release under invalidated approvals, while
reevaluation requires recorded progress within each transaction's retry
budget. Building on this contract, we formulate evidence allocation to
maximize the expected whole-period rate of timely valid releases under
risk, quality-of-service, deadline, and shared byte-budget constraints.
For energy-saving and slice-protection workloads with stable daily load
cycles, thresholds are selected offline and interpolated using an online
load-phase estimate. On average over modeled start phases and quota
levels, the allocation policy improves timely valid release rates over
arrival-order fetching and an offline-selected global threshold under
identical quotas and release checks. In OpenAirInterface radio
emulation, an immediate scripted resend after a version-mismatch refusal
raises on-time receiver delivery from 77.3\% to 89.9\% at modest
additional command cost.

%% file: sections/1_introduction.tex
\section{Introduction}\label{sec:introduction}

\begin{figure*}[t]
    \centering
    \includegraphics[width=\textwidth]{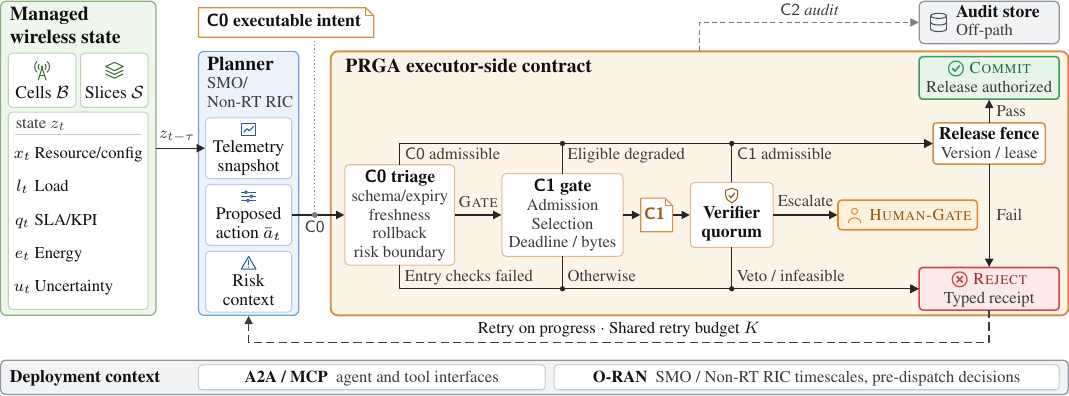}
    \caption{PRGA's execution flow links selective evidence
    acquisition, version-bound release and bounded resubmission.}
    \label{fig:architecture}
\end{figure*}

Agentic artificial intelligence (AI) built on large language models
(LLMs) interleaves reasoning with environment-facing
actions~\cite{yao2023react,wang2024llmagents}, and planner agents that
formulate control intents for actuation are being explored for network
automation~\cite{wang2026comagent,elkael2025agentran,li2026multiagentic}.
In the Open Radio Access Network (O-RAN) architecture, supervisory
control decisions are made at the Service Management and Orchestration
(SMO) and Non-Real-Time RAN Intelligent Controller (Non-RT~RIC) layers
on a seconds-to-minutes timescale~\cite{polese2023oran,oran2024wg2};
typical use cases are energy saving (UC1) at low load and slice
service-level-agreement (SLA) protection (UC2) when traffic,
interference, or radio conditions shift.
At the executor boundary, however, an AI-generated intent is no longer
only a high-level recommendation: it may become a concrete supervisory
operation, such as putting a cell into sleep mode, reducing RF
power, changing slice priority, restricting admission or reallocating
radio resources. Reliable network automation therefore requires an
executor that can assess whether each proposed operation has sufficient
evidence and remains admissible at release.

Consider a planner that observes a low-load snapshot and proposes
putting a cell into sleep mode. Whether the sleep is safe depends on
conditions that the intent does not report, such as the capacity of
neighboring cells; the executor learns them from coordination evidence.
While that evidence is fetched, another planner's slice-protection
update may change the configuration of overlapping resources, so the
sleep's approval must be rechecked before release. If the recheck fails,
refusing the release protects the network but leaves the control
objective unresolved, and the planner may resubmit.

The executor is the last accountable boundary before an intent is
released for network actuation. Acquiring evidence consumes time before
the deadline and, under a shared quota, bytes that later requests cannot
use. An approval can lose its validity when the resource versions or
lease change, even if its evidence remains recent. A refused intent may
also return, so reevaluation needs a progress condition and a bound on
evaluations per transaction. Prior work provides mechanisms for runtime
assurance~\cite{simplex2023,lv2024safemarl},
shielding~\cite{alshiekh2018shielding,dong2026shieldric}, pre-actuation
contract audit~\cite{bimo2026caif}, and runtime
enforcement~\cite{wang2026agentspec}. These
mechanisms provide safeguards at particular stages of execution, but a
supervisory executor must also coordinate them across the lifetime of an
intent. The challenge is to allocate limited evidence among competing
requests while rechecking approvals at release and keeping reevaluation
within each transaction's retry budget.

To address this challenge, we propose Progressive Risk-Gated
Actuation (PRGA), an executor-side actuation model and contract for
agentic wireless supervisory control (Fig.~\ref{fig:architecture}). The
planner supplies candidate intents, and the executor controls their
evidence acquisition and authorization for release. Executable intent
fields ($\Czero$) support local checks, while coordination evidence
($\Cone$) is retrieved on selected gated paths. A provenance digest
($\Ctwo$) supports auditing outside the online safety path. Our
contributions are threefold.

\smallskip\noindent\textbf{1) An executor-side actuation model and contract.}\quad
 We develop an executor-side model and contract to govern the
transition from agent-generated intents to network actuation. The
contract preserves mandatory evidence requirements under resource
shortages, rechecks approvals against bound versions and leases before
release, and permits reevaluation only on receipt-defined progress
within the parent transaction's retry budget. Under stated enforcement
assumptions, it prevents bypass of the specified hard checks, bounds
evaluations per parent transaction, and records every terminal
decision.

\smallskip\noindent\textbf{2) Evidence allocation under a shared quota.}\quad
 To allocate limited evidence across competing requests, we
formulate pre- and post-evidence decisions under risk,
quality-of-service (QoS), deadline, and byte constraints. For
\mbox{quota-limited} \mbox{operation}, thresholds are selected offline from a finite
set to maximize whole-period timely valid releases under a risk screen,
then interpolated using an online load-phase estimate. In the modeled
workloads, averaged over start phases, this policy improves timely valid
release rates over arrival-order fetching and an offline-selected global
threshold under the same quotas and release checks.

\smallskip\noindent\textbf{3) Evaluation of operational effects and costs.}\quad
 To assess PRGA's operational value, we combine comparisons with
reference executors implemented from independent designs,
OpenAirInterface (OAI) radio emulation, and offline experiments on
resolution after rejection. PRGA's deterministic instance is compared
with these executors under a common risk cap, and the emulation
demonstrates reservation-based release protection. With a scripted controller, immediate resend after a
version-mismatch refusal raises on-time receiver delivery from $77.3\%$
to $89.9\%$ at additional command cost. The offline experiments show that bounded resolution
terminates every evaluated parent transaction within its retry budget.

%% file: sections/2_related_work.tex
\section{Related Work}\label{sec:related_work}

\begin{table*}[!t]
\centering
\caption{Reported capabilities of related work and PRGA at the
supervisory executor boundary.}
\label{tab:comparison}
\renewcommand{\arraystretch}{1.0}
\setlength{\tabcolsep}{3pt}
\footnotesize
\begin{tabular}{l l c c c c c c}
\toprule
\textbf{Work}
  & \textbf{Domain}
  & \textbf{\shortstack{Exec.\ Txn.\\Semantics}}
  & \textbf{\shortstack{Sel.\ Evidence\\Acq.}}
  & \textbf{\shortstack{Freshness \&\\Version Binding}}
  & \textbf{\shortstack{Bounded\\Repair}}
  & \textbf{\shortstack{Resource-Aware\\Sel.}}
  & \textbf{\shortstack{External\\RAN Eval.}} \\
\midrule
CAIF \cite{bimo2026caif}
  & O-RAN
  & \pmark & \xmark & \xmark & \xmark & \xmark
  & \cmark \\
SHIELD-RIC \cite{dong2026shieldric}
  & O-RAN
  & \pmark & \xmark & \xmark & \xmark & \xmark
  & \xmark \\
AgentRAN \cite{elkael2025agentran}
  & Wireless
  & \pmark & \pmark & \xmark & \xmark & \xmark
  & \cmark \\
Li et~al.\ \cite{li2026multiagentic}
  & Wireless
  & \pmark & \xmark & \xmark & \xmark & \xmark
  & \xmark \\
O-RAN A1 \cite{polese2023oran,oran2024wg2}
  & Wireless
  & \pmark & \pmark & \pmark & \xmark & \xmark
  & \xmark \\
IBN \cite{clemm2022rfc9315,leivadeas2023ibn,njah2025ibn}
  & Networking
  & \pmark & \xmark & \xmark & \pmark & \xmark
  & \xmark \\
Simplex \cite{simplex2023}
  & CPS / Safety
  & \xmark & \xmark & \xmark & \xmark & \xmark
  & \xmark \\
Shielding / runtime enforcement \cite{alshiekh2018shielding,lv2024safemarl,wang2026agentspec}
  & RL / agents
  & \pmark & \xmark & \xmark & \pmark & \xmark
  & \xmark \\
\midrule
\textbf{PRGA}
  & \textbf{Wireless}
  & \cmark & \cmark & \cmark & \cmark
  & \cmark & \pmark \\
\bottomrule
\end{tabular}

\vspace{2pt}
\begin{minipage}{\textwidth}
\footnotesize
Reported support: \cmark{} native, \pmark{} partial, \xmark{} not
reported in the cited source. PRGA's external RAN evaluation covers
release protection and recovery on the OAI radio-frequency simulator
(RFsim). Evidence selection is evaluated offline.
\end{minipage}
\end{table*}

\Cref{tab:comparison} compares whether a design defines
commit/gate/reject transitions for translated intents, acquires evidence
in response to a gate reason, and terminates stale decisions while
binding release to the evaluated resource generation. It also records
reason-carrying repair with bounded rounds, evidence selection under
explicit deadline and byte budgets, and evaluation against an external
RAN stack.

\subsection{Agentic AI and Multi-Agent Wireless Control}
\label{ssec:rw_agentic_wireless}

AgentRAN~\cite{elkael2025agentran} deploys LLM-powered agents across
O-RAN timescales with over-the-air validation, and
Com\-Agent~\cite{wang2026comagent} uses a multi-LLM
perception--planning--action--reflection loop for wireless optimization.
Li \emph{et~al.}~\cite{li2026multiagentic} introduce a
perception/reasoning/refinement stack with retrieval-augmented reasoning
for conflict-aware \emph{rApp} policy orchestration in Open~RAN.
AutoGen~\cite{wu2024autogen} coordinates workflows across heterogeneous
agents. One level upstream of intent dispatch, large language
models have been used to solve the wireless optimization problems
themselves, for example multi-objective optimization for integrated sensing and
communications (ISAC) in unmanned aerial vehicle (UAV)
networks~\cite{li2025llmisacuav} and joint user
association and beamforming design for ISAC
networks~\cite{li2025llmisacbeam}. These methods improve the quality of the
proposal, whereas PRGA governs whether a proposal may be released to the
network at all.
Semantic-communication
work~\cite{qin2023semcom,yang2023semcom} optimizes how task-relevant
meaning is encoded over noisy links. These works address planner
reasoning, agentic control architectures and PHY/link-layer encoding,
and at most part of PRGA's commit/gate/reject semantics.

\subsection{O-RAN, IBN, and Supervisory Policy Control}
\label{ssec:rw_oran_ibn}

The O-RAN architecture~\cite{polese2023oran} provides a supervisory
control plane through the SMO and Non-RT~RIC, with the A1
interface~\cite{oran2024wg2,etsi2025a1gap,etsi2025a1ap} supporting
policy and enrichment-information exchange on seconds-to-minutes
timescales. Intent-based networking (IBN) formalizes the intent lifecycle --- specification,
translation, activation, assurance --- as surveyed by Leivadeas and
Falkner~\cite{leivadeas2023ibn}, described in IRTF Informational
RFC~9315~\cite{clemm2022rfc9315}, extended by Mehmood
\emph{et~al.}~\cite{mehmood2023ibn} for cellular networks, and given
AI-driven refinement by Njah \emph{et~al.}~\cite{njah2025ibn}. These
works define the placement, interfaces and lifecycle governing what the
network should achieve; PRGA operates inside the activation boundary
\emph{after} translation, deciding commit/gate/reject under freshness,
rollback and risk-divergence constraints.

\subsection{Runtime Assurance, Pre-Actuation Contract Audit, and Safe Actuation}
\label{ssec:rw_runtime_assurance}

Runtime assurance (RTA) focuses on safety monitors, switching logic, and
fault handling. Dynamic Simplex~\cite{simplex2023} pairs an
unverified high-performance controller with a verified safety controller
and switches between them pre-actuation when the runtime monitor
predicts loss of recoverability. Shielding for safe reinforcement
learning~\cite{alshiekh2018shielding} and classical supervisory control
of discrete-event systems~\cite{ramadge1987supervisory} provide
complementary switching lineages. Safe multi-agent reinforcement
learning~\cite{lv2024safemarl} addresses adversarial communication.
Runtime-enforcement layers for tool-using LLM agents interpose
customizable rules on each proposed
action~\cite{wang2026agentspec}. Each of these evaluates the
\emph{action}: a shield constrains or projects a candidate against a
safety specification, and Simplex/RTA substitutes a verified controller
before the next action is applied. PRGA instead decides whether a
translated supervisory transaction carries sufficient, fresh and
version-consistent \emph{evidence} for release, selectively acquires
what is missing under a deadline and a byte budget, and returns
structured reject/repair semantics.

Two recent O-RAN designs place a deterministic check immediately before
actuation and are the closest related work.
\emph{SHIELD-RIC}~\cite{dong2026shieldric} couples an LLM planner
proposing cross-layer recovery actions for a self-healing Open~RAN edge
with a deterministic shield filtering every candidate against a fixed
predicate set before it is applied. In the reported aggregate
simulation results, the shielded arm records no predicate violations
in any scenario, at a packet-delivery-ratio cost. That is compliance
with the deployed predicate set; its shield is a per-action admission
filter, not an evidence-acquisition policy.
\emph{CAIF}~\cite{bimo2026caif} audits a translated user objective
against a formal RAN contract before actuation and validates the
resulting slicing policy on an O-RAN testbed; its chain runs semantic
intent $\rightarrow$ formal contract $\rightarrow$ policy generation and
validation, whereas PRGA's runs translated supervisory transaction
$\rightarrow$ evidence sufficiency and admission $\rightarrow$
conditional downstream release. Both impose a deterministic guardrail
before an AI-originated intent reaches the network and treat a contract
violation as a hard stop. PRGA additionally specifies how evidence obligations,
version-bound release and bounded resolution constrain one another
within a supervisory transaction. Its gate also
conditions on telemetry staleness, rollback validity and
planner--executor risk divergence, and it acquires $\Cone$ selectively
by gate reason under an explicit deadline and byte budget.

\subsection{Agent Interoperability, Tool Invocation, and Provenance}
\label{ssec:rw_a2a_mcp_provenance}

\leavevmode The Agent2Agent (A2A) protocol~\cite{a2a2025} defines agent
discovery, task lifecycles and exchanges between agents. The Model
Context Protocol (MCP)~\cite{mcp2025} standardizes tool invocation and
access to resources and prompts. Both include provisions for
authentication and access control. Permission to invoke an agent or
tool does not establish that a proposed wireless action remains
admissible at dispatch. PRGA complements these interfaces with execution
rules for evidence acquisition, version-bound release and bounded
reevaluation after refusal. Work on Trust, Risk, and Security Management
for agentic AI~\cite{raza2026trism} addresses broader governance and
accountability. PRGA's $\Ctwo$ digest supports this accountability role
after the decision, while PRGA's admission checks and release fence
govern online authorization.

%% file: sections/3_system_model.tex
\section{Wireless Supervisory Network Model and Problem Formulation}
\label{sec:system_model}

\subsection{Wireless Supervisory Network State}
\label{ssec:network_state}

The supervisory loop runs on a seconds-to-minutes timescale over finite
sets of managed radio nodes $\mathcal{B}$ (cells / base stations) and
services $\mathcal{S}$ (slices / SLA-bearing services). At decision
epoch $t$ the network state is $z_t = (x_t, l_t, q_t, e_t, u_t)$, where
$x_t$ is the controllable resource and configuration vector across
$\mathcal{B}\times\mathcal{S}$ (physical-resource-block (PRB)
allocation, RF power or sleep state,
slice priority class, admission-control parameters), $l_t$ the per-cell
load and per-slice demand vector, $q_t$ the SLA and
key-performance-indicator (KPI) monitoring vector,
$e_t$ the energy state or cost component, and $u_t$ a structured
uncertainty bundle aggregating staleness, conflict intensity,
reversibility class, blocking preconditions, and the planner--executor
risk divergence relevant to the candidate intent. A
benchmark-grounded admissible region $\mathcal{X}_{\mathrm{safe}}$ over
(state, action) pairs is induced by capacity, SLA, slice-isolation,
fairness, staleness, and rollback-feasibility predicates; it is
evaluation truth used to label outcomes after the fact, never a
predicate the executor inspects online. The supervisory transition
$z_{t+1} = F(z_t, \bar a_t, \omega_t)$, with $\bar a_t$ the applied wireless
actuation and $\omega_t$ an exogenous disturbance, is an abstraction for
executor-side reasoning and deterministic benchmark replay. Which components the
benchmark realizes, and which are modeling devices, is stated in
\Cref{ssec:design_goals}.

\subsection{Agentic Planner--Executor Actuation Loop}
\label{ssec:control_loop}

The loop comprises four logical roles: a \emph{planner agent} emitting
candidate control intents, an \emph{executor} that validates each
intent and authorizes its release, \emph{verifier agents} assessing
cross-domain constraints when coordination evidence is required, and an
\emph{audit store} recording post-hoc provenance off the online decision
path. In an O-RAN deployment the planner sits in the SMO or
Non-RT~RIC~\cite{polese2023oran,oran2024wg2} and the executor
authorizes each action's release to the downstream A1 or O1 interface
(\Cref{ssec:a1_pattern}).

At epoch $t$ the planner reasons over a possibly stale snapshot
$z_{t-\tau}$ and emits a candidate intent message $c$ naming a wireless
action from the use-case catalogs of \Cref{sec:profile_policy}. The
executor acts on its online context $\chi_t$ (the $\Czero$ payload and
the executor state, which includes a local estimate of $z_t$): from
$\chi_t$ it computes the transaction state $\statetuple$ of
\Cref{ssec:transaction_state} and decides whether to commit, gate or
reject. Verifiers are consulted only when evidence is fetched.
A message can also carry the common transaction envelope used for local
schema, expiry and idempotency checks (transaction identifier, state
epoch, expiry, idempotency key, visibility scope).

\smallskip\noindent\textbf{Two action levels.}\quad
We distinguish the underlying wireless actuation type
$\bar a_t \in A^{\mathrm{net}}$ (from the typed UC1/UC2 catalog)
from the executor's retrieval and actuation
decision $\eta_t \in \{\,\text{commit}@\Czero,\;\text{fetch-}\Cone,\;
\text{human-gate},\;\text{reject}\,\}$. Replanning after $\REJECT$ is a
planner-side follow-up, not a fifth element of the decision set.
The safety predicates and the transition $F$ depend on $\bar a_t$;
the objective of Problem~P1 also depends on $\eta_t$.

\subsection{Executor-Observable Transaction State}
\label{ssec:transaction_state}

For a candidate intent at epoch $t$, the retrieval policy operates over
a compact executor-observable transaction state
$\statetuple = (\riskt, \deadlinet, \bwt, \conflictt, \staleness,
\reversibility)$: $\riskt \in [0,1]$ the planner-provided risk score,
$\deadlinet > 0$ the remaining time budget before expiry, $\bwt {}\geq 0$ the
available control-plane byte budget at the executor (an
evidence-fetch allowance in bytes, not a link rate), $\conflictt \in [0,1]$
the conflict intensity from the active-intent registry, $\staleness \geq
0$ the age in seconds of the planner's snapshot at the executor's
current epoch, and $\reversibility$ the reversibility class.

The transaction state $\statetuple$ is \emph{not} the full network
state $z_t$: $\deadlinet$ and $\bwt$ are decision-context resource
budgets rather than network quantities, and $u_t$
is candidate-relative, $\staleness$ and $\conflictt$ being computed
against the network state and the active-intent registry while
reversibility, blocking preconditions and risk divergence are evaluated
against the candidate intent itself.

\subsection{Constrained Decision Problem}
\label{ssec:objective_scope}

We formulate the executor's decision as a constrained problem. The
contract defines the feasible modes, and Problem~P1 below states the
objective of evidence allocation under a shared quota. Superscripts
$(0)$ and $(1)$
index the information before and after $\Cone$ retrieval; the decision $\eta_t$ of
\Cref{ssec:control_loop} is the pre-evidence mode $m^{(0)} \in M^{(0)}$ and
the post-evidence mode is $m^{(1)} \in M^{(1)}$, where
\begin{equation}
\label{eq:mode_sets}
\begin{aligned}
M^{(0)} &= \{\,\text{commit}@\Czero,\ \text{fetch-}\Cone,\ \REJECT,\\[-2pt]
        &\qquad\ \ \HUMANGATE\,\},\\
M^{(1)} &= \{\,\text{commit}@\Cone,\ \REJECT,\ \HUMANGATE\,\}.
\end{aligned}
\end{equation}
The pre-evidence decision covers every choice made before $\Cone$
arrives (triage, fetch admission, evidence selection and any degraded-mode
rule), so on the executor of
Algorithms~\ref{alg:c0_triage}--\ref{alg:gate_resolve} it spans
Algorithm~\ref{alg:c0_triage} and part of Algorithm~\ref{alg:gate_resolve};
the post-evidence decision follows $\Cone$ verification. The contract enforces the specified admission and release checks;
the objective is optimized only over invariant-admissible, risk- and
QoS-bounded modes.

\smallskip\noindent\textbf{Hard feasible mask.}\quad
$A^{\mathrm{INV}}_t \subseteq M^{(0)}$ is the invariant-admissible mode
set at epoch $t$, induced by the hard gates of \Cref{sec:profile_policy}
(INV-1--INV-3) together with schema validity, any configured local
reject floor, discharge of blocking preconditions, and the deadline/expiry
margin.
INV-4 separates audit from admission; INV-5 is a final guard checked
after authorization, immediately before release, not a pre-evidence
prediction of version consistency. Requests failing terminal entry
checks are rejected before mode selection; satisfying the resource
certificate does not reopen them. The local reject floor is a
configuration choice, not an invariant.
Algorithms~\mbox{\ref{alg:c0_triage}--\ref{alg:gate_resolve}}
instantiate these conditions for \textsc{PRGA-Fixed};
\Cref{sec:online_selection} gives the quota executor's admission and
release rules.

\smallskip\noindent\textbf{Three-valued QoS.}\quad
Let $Q(a)$ be the QoS constraints of action $a$ (capacity, slice floor,
SLA, fairness). For each $j \in Q(a)$, $\Czero$ triage returns a status in
$\{\mathrm{PASS},\mathrm{FAIL},\mathrm{UNKNOWN}\}$: a $\mathrm{FAIL}$ is
terminal and yields $\REJECT$, while an $\mathrm{UNKNOWN}$ carries
$\bar g^{(0)}_j = +\infty$,
removing commit@$\Czero$ from the feasible set and routing the epoch to
the evidence-mandatory class. A verifier emits a \emph{quantitative}
margin $\bar g_j = \hat g_j + m_j$, where $\hat g_j$ is the predicted
signed constraint residual and $m_j$ is a conservative margin. The
residual is the predicted KPI minus the limit for an upper bound, and
the limit minus the predicted KPI for a lower bound. Approval requires
$\bar g_j \leq 0$. The use-case verifiers of
\Cref{ssec:verifier} are these post-evidence evaluators.

\smallskip\noindent\textbf{Feasible set and certificate.}\quad
Writing $\bar p^{(0)}(a \mid \chi_t)$ for the conservative envelope
value of \Cref{ssec:risk_inputs} on $\chi_t$, and
$\varepsilon_a(\mathrm{class}(a))$ for the per-action-class design-time
risk budget, the pre-evidence feasible set of a surviving request is
mode-conditioned --- each governed mode enters through its own
predicate, the terminal modes unconditionally:
\begin{equation}
\label{eq:feasible_set}
\begin{aligned}
\mathcal{F}^{(0)}(\chi_t) ={}& \{\REJECT,\ \HUMANGATE\}
  \cup \bigl\{\text{fetch-}\Cone : \mathrm{Cert}_t\bigr\}\\
 &\cup \bigl\{\text{commit@}\Czero(a) \in A^{\mathrm{INV}}_t :\
    \bar p^{(0)} \leq \varepsilon_a,\\[-2pt]
 &\hspace{31mm}\text{QoS-feasible}\bigr\},
\end{aligned}
\end{equation}
so failed predicates can never mask out $\REJECT$ or $\HUMANGATE$, and
every surviving request has a feasible mode. The
resource-feasibility certificate $\mathrm{Cert}_t$ consists of two
checks:
\begin{equation}
\label{eq:certificate}
\widehat{Q}_{1-\varepsilon_f}\!\left(L_{\Cone} \mid \chi_t\right) + L_{\mathrm{fence}} \leq \deadlinet,
\quad
\widehat{B}_{\Cone}(\chi_t) \leq \bwt .
\end{equation}
The latency check compares a conditional latency quantile plus the
fence revalidation reserve with the remaining deadline. The byte check
compares a deterministic schema-cap estimate of the serialized $\Cone$
size with the per-intent allowance. Here, $L_{\Cone}$ is the $\Cone$
transfer-plus-verification latency, and
$1-\varepsilon_f$ is the chosen latency-quantile level;
Table~\ref{tab:parameter_provenance} specifies the estimate-based
implementation, which uses the $0.95$ verifier-delay quantile. The
certificate is an \emph{estimated} criterion; it neither excludes a
deadline miss nor bounds the miss rate. $L_{\mathrm{fence}}$ is a
configured reserve: the quota runs of
\Cref{ssec:rq5_quota} set it to the pooled fence execution-time p99
measured in \Cref{ssec:rfsim_impl}, and the other runs reserve none.

\smallskip\noindent\textbf{Two branches.}\quad
If commit@$\Czero \in \mathcal{F}^{(0)}(\chi_t)$, the request may commit
without evidence; an evidence-mandatory reason or unknown QoS
removes commit@$\Czero$, so an allocation rule that withholds the fetch
leaves only $\REJECT$ or $\HUMANGATE$. The residual $\MRISK$
gate also masks ordinary commit@$\Czero$ but is not evidence-mandatory:
its guarded degraded refinement is in Algorithm~\ref{alg:gate_resolve}.

\smallskip\noindent\textbf{Problem P1.}\quad
A budget period of $T$ epochs holds $T B_{\mathrm{budget}}$ bytes of
$\Cone$ evidence. A \emph{timely valid release} passes three checks
when it is released: its deadline has not passed, its bound resource
generation is unchanged, and its state age is within the staleness bound $\deltastale$. Over
policies $\pi$ mapping the online context to a feasible mode, the design
problem is
\begin{equation}
\label{eq:p1}
\max_{\pi}\ \mathbb{E}\Bigl[\,\frac{1}{T}\sum_{t=1}^{T} V_t\Bigr]
\quad\text{s.t.}\quad m^{(0)}_t \in \mathcal{F}^{(0)}(\chi_t),
\end{equation}
the post-evidence feasibility conditions of \Cref{ssec:stage2}, and the
period budget
\begin{equation}
\label{eq:byte_budget}
\sum_{t=1}^{T} B^{\Cone}_t \;\leq\; T\,B_{\mathrm{budget}}
\quad [\text{bytes}],
\end{equation}
where $V_t=1$ if the intent of epoch $t$ ends in a timely valid release
and $V_t=0$ otherwise, and $B^{\Cone}_t$ is its settled $\Cone$ bytes,
zero without a fetch. P1 measures the executor's ability to serve
planner-proposed intents under a shared evidence budget by assigning
unit value to each timely valid release. The use-case KPIs separately
assess the resulting network effects. Safety enters through the feasible set: every commit
must satisfy the class-specific envelope bound $\varepsilon_a$, which
evidence allocation cannot relax. The quota
executor of \Cref{ssec:prga_ps} enforces~\eqref{eq:byte_budget} by
reservation and settlement under any selection rule, and
Problem~\eqref{eq:q} selects evidence thresholds for P1 from a declared
finite family.

We label outcomes against use-case-specific safety predicates over the
committed action $\bar a_t$ and the ground-truth scenario state:
$\phi_{\mathrm{UC1}}$ holds when post-action serving-cell load remains
within capacity and the user-throughput change stays within the SLA
limit on the affected cells in $\mathcal{B}$; $\phi_{\mathrm{UC2}}$
holds when the post-action slice-SLA violation rate remains within the
contracted threshold on every affected slice in $\mathcal{S}$. Both are
scored from scenario truth, never seen by the executor online.

\subsection{Information Structure}
\label{ssec:information_structure}

We distinguish three kinds of information.
\emph{Evaluation truth} is scenario ground truth (the per-candidate
unsafe labels), assigned after the fact, never seen online.

\emph{Online proxies} are the quantities computable before actuation,
and they are the only ones the decision may read, among them the planner
risk score $\riskt$, the conservative envelope values $\bar p^{(0)}$ and
$\bar p^{(1)}$, and the predicted QoS constraint residuals $\hat g$
and their margin-added counterparts $\bar g$.

\emph{Delayed outcomes} are observable only for executed actions, hence
selective and censored; they enter evaluation, never the online
decision.

Two disciplines follow. Risk labels are not drawn from committed samples
only, which would bias calibration toward the decisions the policy
already makes; and the unsafe-label predicate is computed from
post-decision state variables that are never inputs of the risk model,
so the model is not fitted to a target encoding its own output.

\subsection{Threat Model, Design Goals, and Benchmark Correspondence}
\label{ssec:threat_model}\label{ssec:design_goals}

Five failure modes shape executor-side actuation under stale and
conflicting state: \emph{stale telemetry}, \emph{conflicting intents}
from concurrent planner threads~\cite{adamczyk2023oranconflict},
\emph{deadline pressure}, \emph{missing or invalid rollback handles},
and \emph{missing provenance}. The contract of
\Cref{sec:profile_policy} addresses the first four through its execution
rules. The $\Ctwo$ digest supports provenance after the online
decision. The threat model is non-adversarial: no planner, verifier
or other controller acts maliciously. PRGA aims to reduce executor
decision latency and online retrieval bytes through local triage when
safe, while preserving the contract's invariants and fetching evidence
only on gated paths within deadline and byte budgets. The contract also
governs authorization validity at release and bounds reevaluation after
rejection.

\smallskip\noindent\textbf{Model-to-benchmark correspondence.}\quad
 Each model element is fully realized by the benchmark, partially
realized, or a modeling device connecting executor checks to evaluation
truth. \emph{Fully realized}: $q_t$; the action pair
$\bar a_t,\eta_t$. \emph{Partially realized}: the safety predicates
$\phi_{\mathrm{UC1}}/\phi_{\mathrm{UC2}}$ (labels drawn from a load- and
action-dependent unsafe probability); the registries
$\mathcal{B},\mathcal{S}$; $x_t, l_t, e_t$ ($e_t$ UC1-only; $l_t$ from
the scenario engine, not an external 3GPP load model); $u_t$;
$\statetuple$ with its resource budgets; the disturbance $\omega_t$; the
post-hoc $\Ctwo$ digest. \emph{Modeling devices}: $z_t$,
$\mathcal{X}_{\mathrm{safe}}$, and the replay transition $F$. Scenario
context follows TR~38.864~\cite{3gpp38864} for UC1 and
TS~28.541~\cite{3gpp28541}, with KPI templates from
TS~28.554~\cite{3gpp28554}, for UC2.

%% file: sections/4_profile_policy.tex
\section{Progressive Control-Intent Actuation Contract}
\label{sec:profile_policy}

PRGA maps an incoming control intent to one of $\COMMIT$, $\GATE$, or
$\REJECT$ before any wireless action is applied. Its online path
uses $\Czero$ and, when fetch admission permits, $\Cone$.

\smallskip\noindent\textbf{$\COMMIT$ semantics.}\quad
Throughout, $\COMMIT$ means \emph{commit-to-downstream}: the executor's
authorization to release the intent, which the contract governs.
Transmission of the downstream request (an A1-P Policy Create/Update
or an O1 configuration change, \Cref{ssec:a1_pattern}), its acceptance by the downstream controller,
which can independently reject, and execution in the RAN are later
steps.

Algorithms~\ref{alg:c0_triage}--\ref{alg:gate_resolve} specify the
threshold contract used by PRGA's deterministic instance \textsc{PRGA-Fixed}, including its calibrated
envelope checks.

\subsection{\texorpdfstring{$\Czero$}{C0} --- Executable Control Intent}
\label{ssec:c0}

The $\Czero$ layer is the \emph{minimum executable payload} required for
immediate triage. It can be carried in the common transaction envelope
of \Cref{ssec:control_loop}; the commit/gate/reject semantics below
belong to the contract, not to the envelope.
Representative fields appear in Table~\ref{tab:c0_fields}(a).

\begin{table*}[t]
\centering
\caption{Control-intent specification and default contract settings.}
\label{tab:c0_fields}\label{tab:action_catalog}\label{tab:thresholds}
\footnotesize
\setlength{\tabcolsep}{3pt}
\begin{minipage}[t]{0.295\textwidth}
\centering
\begin{tabular}[t]{@{}>{\raggedright\arraybackslash\leavevmode}p{2.55cm}>{\raggedright\arraybackslash\leavevmode}p{2.55cm}@{}}
\multicolumn{2}{@{}l}{\textbf{(a) Representative $\Czero$ fields}}\\
\toprule
\textbf{Field} & \textbf{Description} \\
\midrule
\texttt{intent\_type}, \texttt{proposed\_action} & Catalog action type, cf.\ (b); parameterized spec. \\
\texttt{target\_scope}, \texttt{resource\_keys}  & Managed objects; affected resources \\
\texttt{state\_epoch}, \texttt{expires\_at}      & Snapshot epoch; absolute expiry \\
\texttt{reversibility\_\allowbreak class}        & \{reversible, costly-rev., irreversible\} \\
\texttt{risk\_score}, \texttt{rollback\_handle}  & Planner risk $\in[0,1]$; rollback reference \\
\texttt{needs\_upgrade}, \texttt{blocking\_req}  & Planner $\Cone$ assertion; commit prerequisites \\
\bottomrule
\end{tabular}
\end{minipage}\hfill
\begin{minipage}[t]{0.38\textwidth}
\centering
\begin{tabular}[t]{@{}llccc@{}}
\multicolumn{5}{@{}l}{\textbf{(b) Typed action catalog}}\\
\toprule
\textbf{UC} & \textbf{Action Type} & \textbf{Rev.} & \textbf{Risk} & \textbf{Rlz.} \\
\midrule
\multirow{5}{*}{\rotatebox[origin=c]{90}{\scriptsize UC1}}
  & \textsc{Cell\_Sleep}       & rev.    & med  & O1 \\
  & \textsc{Cell\_Wake}        & rev.    & low  & O1 \\
  & \textsc{RF\_Power\_Reduce} & rev.    & low  & O1 \\
  & \textsc{RF\_Reconfig}      & costly  & high & O1 \\
  & \textsc{Load\_Redirect}    & rev.    & med  & A1 \\
\midrule
\multirow{5}{*}{\rotatebox[origin=c]{90}{\scriptsize UC2}}
  & \textsc{Slice\_Priority\_Boost}      & rev.    & low  & A1 \\
  & \textsc{Slice\_Admission\_Restrict}  & costly  & med  & O1 \\
  & \textsc{Slice\_Resource\_Realloc}    & costly  & high & A1 \\
  & \textsc{Load\_Balance\_Update}       & rev.    & med  & A1 \\
  & \textsc{SLA\_Escalate}               & irrev.  & low  & model \\
\bottomrule
\end{tabular}
\end{minipage}\hfill
\begin{minipage}[t]{0.305\textwidth}
\centering
\begin{tabular}[t]{@{}lcc>{\raggedright\arraybackslash\leavevmode}p{2.0cm}@{}}
\multicolumn{4}{@{}l}{\textbf{(c) Default threshold parameters}}\\
\toprule
\textbf{Par.} & \textbf{UC1} & \textbf{UC2} & \textbf{Meaning} \\
\midrule
$\taucommit$           & 0.30 & 0.25 & At/below: $\Czero$ risk threshold \\
$\taureject$           & 0.80 & 0.75 & At/above: Stage-1 reject floor \\
$\taudegraded$         & 0.50 & 0.40 & Cap for \MRISK-gated degraded commit \\
$\deltastale$          & 30\,s & 10\,s & Max state age \\
$m_d$            & 5\,s & 2\,s  & Deadline margin at entry \\
$\beta_{\min}$   & 2\,KiB/s & 1\,KiB/s & Minimum link rate for $\Cone$ \\
$\epstrust$            & 0.15 & 0.15 & Risk-divergence tolerance \\
\bottomrule
\end{tabular}
\end{minipage}
\par\smallskip\raggedright
 Rev.: reversibility class. Rlz.: downstream realization (A1 policy, O1
configuration, or model only). Panel (c) gives hand-set defaults fixed before
evaluation.
\end{table*}

\smallskip\noindent\textbf{Typed action catalog.}\quad
Each \texttt{intent\_type} maps to a static tuple of reversibility and
risk class, fixing the safety semantics before any runtime assessment;
Table~\ref{tab:action_catalog}(b) lists the ten action types.
Reversibility determines whether a rollback handle is mandatory (every
class except \emph{irreversible}); the risk class seeds the risk
score. The realization column (Rlz.)\ classifies
each action's downstream path (\Cref{ssec:a1_pattern}).

\subsection{Risk Score and Thresholds}
\label{ssec:risk}

Given a candidate action $a$ and its online context $\chi_t$, the
fixed-threshold executor computes a local composite risk score:
\begin{equation}
\label{eq:risk_score}
r_\ell(a,\chi_t) \;=\; w_{\!t}\,\phi(a)
  \;+\; w_{\!s}\,\hat\sigma(\chi_t)
  \;+\; w_{\!c}\,\hat c(\chi_t)
  \;+\; w_{\!n}\,\hat n(\chi_t),
\end{equation}
where $\phi(a)\!\in\!\{0.2,0.5,0.8\}$ maps the risk class to a numeric
value, $\hat\sigma = \min(|\text{epoch}_{\mathrm{exec}} -
\texttt{state\_epoch}|\,/\,\deltastale,\;1)$ is normalized
staleness~\cite{kaul2012aoi}, $\hat c$ the conflict intensity and
$\hat n$ resource contention (mean target utilization of the intent's resource keys).
We use $(w_{\!t},w_{\!s},w_{\!c},w_{\!n})=(0.3,0.3,0.2,0.2)$ as an
interpretable default, which weights the action's risk class and
staleness slightly above conflict intensity and resource contention.
This vector and the thresholds in Table~\ref{tab:thresholds}(c)
are hand-assigned operator defaults. \textsc{PRGA-Fixed} combines this
score with the calibrated envelope of \Cref{ssec:risk_inputs} as an
additional commit-feasibility requirement.

\subsection{Stage~1: \texorpdfstring{$\Czero$}{C0} Triage}
\label{ssec:stage1}

Algorithm~\ref{alg:c0_triage} specifies the $\Czero$ triage executed
locally by the executor upon receiving a control intent.  Every
commit path in
Algorithms~\ref{alg:c0_triage}--\ref{alg:gate_resolve} begins with this
triage, which is deterministic given the $\Czero$ payload, the current
time, and the executor's local state and configuration.

\begin{algorithm}[t]
\caption{Stage~1: $\Czero$ Contract Triage}
\label{alg:c0_triage}
\footnotesize\raggedright
\begin{algorithmic}[1]
\REQUIRE $\Czero$ intent $c$, current time $t_{\mathrm{now}}$, executor state $\mathcal{E}$
\ENSURE $(\text{decision},\text{gate\_reason})$ with
 $\text{decision}\!\in\!\{\COMMIT,\allowbreak\GATE,\allowbreak\REJECT\}$ and
 $\text{gate\_reason}\!\in\!\{\GRNONE,\allowbreak\RDIV,\allowbreak\LCONFL,\allowbreak\PUPG,\allowbreak\mathrm{QUNK},\allowbreak\MRISK\}$;
 $\text{gate\_reason}\!=\!\GRNONE$ whenever $\text{decision}\!\neq\!\GATE$
\IF{$\neg\,\textsc{SchemaValid}(c)$}
  \RETURN $(\REJECT,\GRNONE)$\;\texttt{// schema invalid}
\ENDIF
\IF{$\textsc{EnvelopeDefined}(c)$ \AND $\neg\,\textsc{ValidEnvelope}(c,\mathcal{E})$}
  \RETURN $(\REJECT,\GRNONE)$\;\texttt{// envelope invalid}
\ENDIF
\IF{$t_{\mathrm{now}} \geq c.\texttt{expires\_at} - m_d$}
  \RETURN $(\REJECT,\GRNONE)$\;\texttt{// expired}
\ENDIF
\IF{$|\mathcal{E}.\text{epoch} - c.\texttt{state\_epoch}| > \deltastale$}
  \RETURN $(\REJECT,\GRNONE)$\;\texttt{// stale (INV-1)}
\ENDIF
\IF{$c.\texttt{reversibility\_class} \neq \text{irreversible}$ \AND $\neg\,\textsc{ValidRollback}(c.\texttt{rollback\_handle},\mathcal{E})$}
  \RETURN $(\REJECT,\GRNONE)$\;\texttt{// rollback invalid (INV-2)}
\ENDIF
\STATE $r_{\ell} \gets \textsc{ComputeRisk}(c,\mathcal{E})$ \COMMENT{Eq.~\eqref{eq:risk_score}}
\IF{$r_{\ell} \geq \taureject$}
  \RETURN $(\REJECT,\GRNONE)$\;\texttt{// local reject floor}
\ENDIF
\STATE $Q_0 \gets \textsc{QoSStatus}(c,\mathcal{E})$
\IF{$\mathrm{FAIL}\in Q_0$}
  \RETURN $(\REJECT,\GRNONE)$\;\texttt{// violated necessary condition}
\ENDIF

\IF{$|c.\texttt{risk\_score} - r_{\ell}| > \epstrust$}
  \RETURN $(\GATE,\RDIV)$\;\texttt{// risk divergence (INV-3)}
\ENDIF
\IF{$c.\texttt{needs\_upgrade}$}
  \RETURN $(\GATE,\PUPG)$\;\texttt{// planner upgrade}
\ENDIF
\IF{$\textsc{LocalConflict}(c,\mathcal{E})$}
  \RETURN $(\GATE,\LCONFL)$\;\texttt{// local conflict}
\ENDIF
\IF{$\mathrm{UNKNOWN}\in Q_0$}
  \RETURN $(\GATE,\mathrm{QUNK})$\;\texttt{// evidence required}
\ENDIF

\IF{$r_{\ell} \leq \taucommit$ \AND $c.\texttt{blocking\_req}=\emptyset$ \AND $\bar p^{(0)}\leq\varepsilon_a$}
  \RETURN $(\COMMIT,\GRNONE)$\;\texttt{// risk decision}
\ELSE
  \RETURN $(\GATE,\MRISK)$ \COMMENT{mid-risk residual}
\ENDIF
\end{algorithmic}
\end{algorithm}

\smallskip\noindent\textbf{Local predicates.}\quad
$\textsc{EnvelopeDefined}(c)$ holds when the message carries the
transaction envelope of \Cref{ssec:control_loop}, and
$\textsc{ValidEnvelope}$ iff that envelope's fields are well-formed and
consistent with the executor's view (vacuously satisfied with no
envelope attached, as in the reported runs). $\textsc{ValidRollback}(h,\mathcal{E})$ holds iff
$h$ is non-null, targets a scope reachable from $\mathcal{E}$, matches
the policy version the intent was issued under, is unexpired, names an
available rollback procedure and has not been consumed. Both are local
and require no $\Cone$. The handle is checked for structural validity at
Stage-1 entry, with the verdict retained through release. This does not
establish physical recoverability.
The offline engine implements only the handle-presence clause; the
other clauses are unmodeled there.

\smallskip\noindent\textbf{Stage-1 output contract.}\quad
 Stage~1 emits a decision and a reason: $\RDIV$, $\PUPG$,
$\LCONFL$ and $\mathrm{QUNK}$ (unknown QoS status) identify their priority-ordered branches;
$\MRISK$ is the residual risk or blocking-precondition case, including
a failed calibrated-envelope check; $\GRNONE$ accompanies a non-gated
decision. The local reject floor terminates
$r_\ell\geq\taureject$ before these gates.

\smallskip\noindent\textbf{Invariants.}\quad
Triage enforces three safety invariants by construction.
\textbf{INV-1}: no intent with
$|\mathcal{E}.\text{epoch}-c.\texttt{state\_epoch}|>\deltastale$
at Stage-1 entry is ever committed. \textbf{INV-2}: no non-irreversible intent without a
$\textsc{ValidRollback}$ handle is ever committed; irreversible intents
are governed by the typed catalog instead. \textbf{INV-3}: $\RDIV$,
$\LCONFL$, $\PUPG$ and $\mathrm{QUNK}$ are evidence-mandatory gates. An
intent behind one of these gates may commit only through
Algorithm~\ref{alg:gate_resolve}'s admitted $\Cone$-fetch branch, with a
verifier-approved quorum and $c.\texttt{blocking\_req}=\emptyset$.
Failed fetch admission yields $\REJECT$. These gates remove
commit@$\Czero$ from the feasible set of \Cref{ssec:objective_scope}.
Risk divergence is checked against a locally recomputed score.

\subsection{Stage~2: \texorpdfstring{$\GATE$}{GATE} Resolution}
\label{ssec:stage2}

Stage-1 $\GATE$ outputs enter Algorithm~\ref{alg:gate_resolve}.
Admitted fetches obtain $\Cone$ and run \textsc{VerifierQuorum}; failed
admission invokes the degraded-mode rule.

\begin{algorithm}[t]
\caption{Stage~2: $\GATE$ Resolution}
\label{alg:gate_resolve}
\footnotesize
\begin{algorithmic}[1]
\REQUIRE Gated intent $c$, gate reason $g$, remaining deadline $d$, link rate $\beta$, optional byte allowance $b_t$, risk $r_{\ell}$
\ENSURE decision $\in \{\COMMIT, \REJECT, \HUMANGATE\}$
\IF{$\textsc{FetchAdmissible}(c,d,\beta,b_t)$}
  \STATE $c_1 \gets \textsc{Fetch\_C1}(c)$
  \STATE \algorithmicif\ $c_1=\bot$ \algorithmicthen\ \algorithmicreturn\ $\REJECT$ \COMMENT{failed or late}
  \STATE $v \gets \textsc{VerifierQuorum}(c, c_1)$ \COMMENT{Sec.~\ref{ssec:verifier}}
  \IF{$v = \text{APPROVED}$}
    \IF{$c.\texttt{blocking\_req} \neq \emptyset$ \OR $\bar p^{(1)}>\varepsilon_a$}
      \RETURN $\REJECT$ \COMMENT{commit infeasible}
    \ENDIF
    \RETURN $\COMMIT$
  \ELSIF{$v = \text{CONFLICT}$}
    \RETURN $\REJECT$
  \ELSE
    \RETURN $\HUMANGATE$ \COMMENT{escalate}
  \ENDIF
\ELSE
  \IF{$\textsc{EvidenceMandatory}(g)$}
    \RETURN $\REJECT$ \COMMENT{degraded mode: evidence-mandatory}
  \ENDIF
  \IF{$c.\texttt{blocking\_req} \neq \emptyset$}
    \RETURN $\REJECT$ \COMMENT{blocking\_req forbids degraded commit}
  \ENDIF
  \IF{$g = \MRISK$ \AND $c.\texttt{reversibility\_class} = \text{reversible}$ \AND $r_{\ell} \leq \taudegraded$ \AND $\bar p^{(0)}\leq\varepsilon_a^{\mathrm{deg}}$}
    \RETURN $\COMMIT$ \COMMENT{degraded commit}
  \ELSE
    \RETURN $\REJECT$ \COMMENT{unsafe under resource constraint}
  \ENDIF
\ENDIF
\end{algorithmic}
\end{algorithm}

$\textsc{FetchAdmissible}$ is the configured admission rule.
Estimate-based admission uses \eqref{eq:certificate} and the link-rate
floor. Simulation look-ahead admission compares the request's pre-drawn
transfer-plus-verification time with $d$ and requires
$\beta\geq\beta_{\min}$; it reads the simulated timing before the fetch.
$\textsc{Fetch\_C1}$ returns $\bot$ if the complete $\Cone$ response, including
the received verifier votes, cannot be obtained within $d-L_{\mathrm{fence}}$; the
request is then rejected, its delivered bytes charged. The look-ahead
admits no fetch that returns $\bot$. The byte allowance, set only in the quota runs, is distinct
from the link rate, and the degraded envelope budget
$\varepsilon_a^{\mathrm{deg}}$ is stricter than $\varepsilon_a$.

\smallskip\noindent\textbf{Execution order.}\quad
 The fixed arm resolves a Stage-1 gate with Algorithm~\ref{alg:gate_resolve},
and the threshold policies of \Cref{ssec:prga_ps} may only
withhold an admitted fetch.
All provisional commits then pass INV-5 immediately before release; a
failed fence returns $\REJECT$ and cannot dispatch.

\subsection{\texorpdfstring{$\Cone$}{C1} Evidence and the Verifier Model}
\label{ssec:c1}\label{ssec:verifier}

The $\Cone$ layer carries evidence consumed by verifiers and the
scheduling logic, \emph{not} by the executor for direct actuation:
\texttt{constraint\_summary}, \texttt{conflict\_candidates},
\texttt{missing\_information}, and \texttt{verifier\_votes}. The payload
is a small addition on top of $\Czero$. In
Algorithms~\ref{alg:c0_triage}--\ref{alg:gate_resolve}, an admitted
$\Cone$ fetch follows a Stage-1 $\GATE$, including $\MRISK$.

Verifiers implement a unanimous-approval safety-veto rule: every
vote received must approve, any single veto rejects, and mixed
non-veto outcomes escalate to $\HUMANGATE$. Given a nonempty vote vector
$\boldsymbol{v}$ (an empty one escalates),
\begin{equation}
\label{eq:quorum}
\textsc{Quorum}(\boldsymbol{v}) \!=
\begin{cases}
\text{APPROVED}  & \text{if } \forall v_i\!: v_i \!=\! \text{APPROVE},\\
\text{CONFLICT}  & \text{if } \exists v_i\!: v_i \!=\! \text{VETO},\\
\text{ESCALATE}  & \text{otherwise.}
\end{cases}
\end{equation}
The domain verifiers are a \emph{Load Verifier} and an \emph{SLA
Verifier} for UC1, and for UC2 a \emph{Slice Isolation Verifier}
(vetoing if a slice's guaranteed minimum resource is violated) and a
\emph{Fairness Verifier} (vetoing a low Jain
index~\cite{jain1984fairness}); the scenario engine replaces them
with the simulated votes of \Cref{ssec:baseline_fidelity}.

\subsection{INV-5: Version-Bound Authorization Fence}
\label{ssec:inv5}

\nobreak
A provisional approval is issued against a particular view of the managed
resources, which can be superseded between approval and dispatch. INV-5
closes that window.

\smallskip\noindent\textbf{INV-5.}\enspace
A provisional approval remains valid only if an executor--adapter
compare-and-set on the relevant resource lease and the expected
policy/configuration generation succeeds immediately before downstream
dispatch; a mismatch invalidates the authorization.

Immediately before dispatch, after risk, QoS and verifier evaluation,
the fence compares only the bound resource keys with their captured
generation, policy version and lease. A mismatch, an unobservable
commit-time generation or an unreachable fence state causes rejection.
The offline engine compares only the bound generation on its commit
path. The coordination registry compares and advances that generation in
the same epoch, and also advances it on lease and policy changes.

In the AgentRAN-integrated OAI RFsim emulation, the adapter performs an
atomic conditional actuation: comparison, actuation and generation
advancement form one atomic step. The adapter also checks the policy
version, when held by the adapter, and any lease constraints installed
on the bound key. A separate reservation guard can refuse a write while
an exclusive lease or another write's reservation overlaps its physical
write set (\Cref{ssec:rq4_rfsim}).

Evidence is verified once, at fetch completion. State freshness is
checked at Stage-1 entry and rechecked at release in the quota runs
(\Cref{ssec:prga_ps}). Other runs retain the entry-time freshness
verdict through the evidence wait, even if the state age then exceeds
$\deltastale$.

\subsection{\texorpdfstring{$\Ctwo$}{C2} --- Provenance Digest}
\label{ssec:c2}

The $\Ctwo$ layer is a post-hoc audit record that is \emph{not} on the
online control path; its content model aligns with W3C
PROV-DM~\cite{w3c2013provdm}, covering telemetry snapshot identifiers,
tool/model/policy/verifier versions, dependency hashes, signature,
evidence URIs and retention class. The contract distinguishes a minimal
decision record, retained on every terminal decision, from the full
$\Ctwo$ digest: Stage-2 and degraded-mode commits emit the full record,
while fast Stage-1 commits need only the minimal one and remain
auditable at that coarser granularity.

\smallskip\noindent\textbf{INV-4 (audit separation).}\quad
 The full $\Ctwo$ digest is generated and stored after the online
decision and does not alter its $\COMMIT$/$\GATE$/$\REJECT$ outcome.

\subsection{Typed Reject Receipt}
\label{ssec:reject_receipt}

A rejection without a machine-readable reason gives the planner little
guidance for repair and can lead to repeated unsuccessful submissions.
Every rejected
executor evaluation therefore returns a typed \emph{Reject Receipt}.
Its core fields identify the attempt (transaction, parent intent,
attempt index and intent fingerprint), state the cause (reason code,
the set of violated predicates and a digest of the evaluated state),
guide the repair (whether and when a retry may follow, a repair hint
and the admissible parameter interval) and mark a rejection that closes
the parent transaction as terminal. It also records the action fingerprint and the bound
resource generation and lease, which deduplication and the progress
predicates compare.

Each reason code carries minimal-repair semantics (the smallest
planner-side change that could make the next attempt admissible) over
stale state, active conflict, invalid rollback handle, risk divergence,
undischarged blocking precondition, insufficient $\Cone$ budget,
verifier veto, and unsatisfiable QoS, derived from the executor's own
terminal path. An empty admissible parameter interval on a quantitative
veto rules out progress through a change of that parameter alone.
Deduplication uses the triple
(reason code, state digest, action fingerprint).

%% file: sections/5_online_selection.tex
\section{Online Evidence Selection and Bounded Resolution}
\label{sec:online_selection}

PRGA separates the executor contract from the policy that allocates
admissible evidence requests.
Algorithms~\ref{alg:c0_triage}--\ref{alg:gate_resolve} specify the
fixed-threshold instance used in the independent-executor and
retrieval-cost comparisons. The quota experiments use a common executor
that enforces the contract and the period budget, while \textsc{FIFO}
(arrival-order fetching), \textsc{GS8} (an offline-selected global threshold), and
\textsc{PRGA-PE-BC} and \textsc{PRGA-PE-BCI} (a phase-estimated threshold
without and with interpolation) differ only in the rule for allocating
evidence. Problem~\eqref{eq:q} selects the
thresholds for this policy family.

The quota executor of
\Cref{ssec:prga_ps} enforces the per-period byte budget by reservation,
capped delivery and settlement, and keeps INV-1--INV-5, but it computes
no score~\eqref{eq:risk_score} and leaves risk feasibility to a fitted
envelope. It has no local reject floor or degraded
commit, and it rejects an undischarged blocking precondition at entry
instead of gating it.

\subsection{Conservative Risk Inputs}
\label{ssec:risk_inputs}

Let $\hat s^{(0)}(\chi_t)$ and $\hat s^{(1)}(\chi_t,e^{\Cone}_t)$ be
the calibrated scores before and after the $\Cone$ evidence
$e^{\Cone}_t$ arrives; their model, pre-specified selection rule and
probability-calibration map are given in \Cref{sec:calibration}.
On a seed-disjoint envelope split, for each action class $a$ and calibrated-score bin $b$ we take the one-sided
Clopper--Pearson upper confidence limit $U_{a,b}$ of the unsafe rate at
a pre-specified level and monotonize it along increasing risk by the
cumulative-maximum majorant
\begin{equation}
\label{eq:majorant}
\bar U_{a,b} \;=\; \max_{j \leq b} U_{a,j},
\qquad
h_a(s) \;=\; \bar U_{a,\,\mathrm{bin}(s)} .
\end{equation}
The cumulative maximum never lowers a per-bin limit. Each bin therefore
retains its one-sided marginal coverage when its observations are
independent and share a common unsafe probability. Dependence among
decisions within a run can violate this condition. The step function $h_a$ is the
final envelope map, and the values consumed online are
\begin{equation}
\label{eq:pbar}
\begin{aligned}
\bar p^{(0)}(a \mid \chi_t) &= h^{(0)}_a\!\bigl(\hat s^{(0)}(\chi_t)\bigr),\\
\bar p^{(1)}(a \mid \chi_t,e^{\Cone}_t) &= h^{(1)}_a\!\bigl(\hat s^{(1)}(\chi_t,e^{\Cone}_t)\bigr).
\end{aligned}
\end{equation}
The calibrated score $\hat s$ never gates a commit by itself: only the
envelope value $\bar p$ enters a feasibility test, and $\varepsilon_a$
is set on that scale. Post-evidence envelope samples
are generated offline for \emph{all} gated candidates, so that envelope
does not depend on the policy's fetches.

\subsection{Quota-Constrained Evidence Selection}
\label{ssec:prga_ps}

\noindent\textbf{Quota enforcement and release metric.}\quad
The per-epoch budget $B_{\mathrm{budget}}$ of \eqref{eq:byte_budget} is
the reference budget scaled by the announced contraction level
$\kappa$, and $\bwt$ in
\eqref{eq:certificate} is the smaller of the per-intent allowance and
the remaining balance. An admitted fetch reserves
$\widehat{B}_{\Cone}(\chi_t)$; the provider delivers at most the
reservation, delivered bytes are settled and the rest refunded, and the
balance is forfeited when the period ends. A delivery truncated at the
reservation is rejected before verification and still charged. Hence
no period settles more than the bytes it holds, whatever
the selection rule, so \eqref{eq:byte_budget} holds. Release re-checks the deadline, the
resource generation and the state age against $\deltastale$, the three
checks of a timely valid release (\Cref{ssec:objective_scope}). Because
bytes settled for one request are unavailable to later requests of the
period, the choice of which requests receive evidence shapes the
whole-period rate of timely valid releases.
The common base, \textsc{FIFO}, takes commit@$\Czero$ when feasible,
else fetch-$\Cone$ when $\mathrm{Cert}_t$ passes, else $\REJECT$, and
drops the fetch for an action class whose
post-evidence envelope exceeds $\varepsilon_a$ in every bin
(\emph{class pruning}); under a binding quota it fetches in arrival
order while the balance covers a reservation. This first-come-first-served
rule is used as a reference in online wireless resource
allocation~\cite{shi2020rlslicing}.\looseness=-1

\smallskip\noindent\textbf{Threshold rule.}\quad
A threshold rule keeps a base fetch whose planner risk score $\riskt$ in $\Czero$ is at
most the threshold $\tau_t$ in force,
\begin{equation}
\label{eq:prga_ps}
\mathrm{fetch}(t) = \mathrm{fetch}_{\mathrm{base}}(t)
\wedge \bigl[\riskt \leq \tau_t\bigr],
\end{equation}
and a request whose fetch is withheld is rejected. In every
executor state, the rule can only withhold a fetch the base would make:
it grants no commit, and every release still passes the same admission
and release checks. It reads online information only: no later
arrival, load function, or pre-drawn $\Cone$ size or verifier delay.
In the model, the score is informative about whether a fetch ends in a
release, since the simulated vote model vetoes more often at a higher
score (\Cref{ssec:baseline_fidelity}). Withholding fetches above a
threshold can therefore leave balance for later requests more likely to
end in a release, but whether it does depends on the deadlines,
versions and arrivals that follow, so Problem~\eqref{eq:q} compares
candidate thresholds by their whole-period closed-loop results.

\smallskip\noindent\textbf{Threshold selection by (Q).}\quad
The global threshold \textsc{GS8} reads one entry per
$(\mathrm{UC},\kappa)$, held over the whole period and fitted jointly over
the eight declared start phases (Table~\ref{tab:parameter_provenance}); it is a
static-threshold policy, a class analyzed for sequential selection under
a capacity limit~\cite{arnosti2023static}. Since in the model the score carries a daily
load term, the best threshold can change with the load phase, so
\textsc{PRGA-PE-BC} and \textsc{PRGA-PE-BCI} below also read
entries indexed by where in the daily load cycle the period starts: a
\emph{phase table} holds one entry
for each declared start phase, and a
\emph{level-class table} one for each \emph{level class}, the declared
start phases of equal starting load. For the phase-aware policies, let
the switching points $n_1<n_2$ be the fixed request counts separating
the global, level-class, and phase-specific segments.

These thresholds are obtained by enumerating a declared finite set
$\Theta$ on closed-loop quota runs of the fitting role, one table entry
$j$ at a time:
\begin{equation}
\label{eq:q}
\tau^{\star}_j\in\operatorname*{arg\,max}_{\tau\in\Theta}\,
\widehat{Y}_j(\tau)
\quad\text{s.t.}\quad \widehat{R}_j(\tau)\le\bar{R}_j .
\tag{Q}
\end{equation}
For each use case, $\Theta$ contains $+\infty$ and quantiles of
base-fetch scores from quota-free fitting runs
(Table~\ref{tab:parameter_provenance}). A candidate $\tau$ is in force
over the whole period for \textsc{GS8}, or from the start of its table's
segment until the period ends for a level-class or phase entry, with
earlier segments at their fitted thresholds. $\widehat{Y}_j(\tau)$ and
$\widehat{R}_j(\tau)$ are the whole-period rates of timely valid
releases and unsafe epochs, averaged over fitting seeds and equally over
$j$'s start phases: all declared phases for \textsc{GS8}, those of its
level class for a level-class entry, and its own phase for a phase
entry. The entry-specific risk screen is
$\bar{R}_j=\max\{\bar{R},\widehat{R}_j(+\infty)\}$, with $\bar{R}$
preset. Ties go to lower spend, then higher $\tau$.
Problem~\eqref{eq:q} thus fits the threshold rules~\eqref{eq:prga_ps}
to P1's release-rate objective on the fitting runs, selecting one table
entry at a time from $\Theta$ under its risk screen.

The tables are fitted in segment order. After fitting \textsc{GS8}, we
hold it fixed and fit the level-class table in a level-only policy that
uses that table after request $n_1$ until the period ends. We then hold
both tables fixed and fit the phase table for the segment after request
$n_2$. Each fitting run contributes to the entry for its true level
class or start phase.

\smallskip\noindent\textbf{Online phase estimation.}\quad
\textsc{PRGA-PE-BC} estimates the starting phase of the modeled load
cycle from previously observed planner risk scores. For each use case
and action class, it uses a fixed periodic mean-score profile fitted
offline. At each switching point, it selects the starting phase that
minimizes the cumulative squared error between the observed scores and
the profile shifted to that phase. The search uses a grid finer than the
declared start phases. Every observed score is included, whether or not
$\Cone$ was fetched, and ties go to the earliest grid phase.

The threshold in~\eqref{eq:prga_ps} is set in three segments:
\begin{enumerate}
\item Requests $1$ through $n_1$ use the \textsc{GS8} threshold.
\item After request $n_1$, the estimate is mapped to the circularly
nearest declared phase, whose level-class entry governs requests
$n_1+1$ through $n_2$.
\item After request $n_2$, a new estimate using all scores observed so
far selects the phase-table entry of its nearest declared phase for the
rest of the period.
\end{enumerate}
Each selected threshold is fixed within its segment. A request's own
score therefore does not enter the phase estimate that sets its
threshold. Fetches in all three segments draw on the same period
balance. Table~\ref{tab:parameter_provenance} gives the grid, switching
counts, and profile construction used in the evaluation.

\smallskip\noindent\textbf{Interpolation.}\quad
\textsc{PRGA-PE-BCI} uses the same first two segments. At the second
switching point, it linearly interpolates between the phase-table
entries at the two declared phases bracketing the estimated grid phase
around the cycle. The resulting threshold serves as $\tau_t$ in
\eqref{eq:prga_ps} until the period ends. Every frozen phase entry is finite, so no interpolation
involves~$+\infty$.

\subsection{Bounded Resolution After Rejection}
\label{ssec:bounded_resolution}

Checks on each executor decision do not by themselves guarantee the
liveness of the planner--executor loop. The property below relies on
the shared retry budget, deduplication, enforced evaluation timeouts
and terminal escalation. Backoff, resource cooldown and the wall-clock
budget can delay resolution or end it early, but cannot increase the
number of executor evaluations beyond the stated bound.

A \emph{parent transaction} is an initial intent together with the
resubmissions made to resolve it. They share one retry budget $K$, a
pre-specified operator parameter. A reevaluation after the bound
resource generation has changed draws on the same budget. A retry is admitted only if the resubmission makes progress (the progress gate), that is,
if at least one of five predicates over the typed receipt of
\Cref{ssec:reject_receipt} holds: the violated-predicate set strictly shrinks;
a named required field is repaired; a quantitative parameter enters the
receipt's admissible interval; the bound resource generation has
changed; or the scheduled retry time has elapsed and the blocking lease
has changed. A
non-progressing resubmission consumes no executor evaluation, and a
second consecutive one trips the circuit breaker, which escalates to
$\HUMANGATE$.

\smallskip\noindent\textbf{Bounded-resolution property.}\enspace
\emph{Under the retry budget $K$, deduplication, the
per-evaluation enforced timeout and terminal escalation, every parent
transaction receives at most $K+1$ executor evaluations, and one whose
planner answers each retryable rejection reaches $\COMMIT$, terminal
$\REJECT$ or $\HUMANGATE$ within them.} Each evaluation is bounded by its
enforced timeout. A progressing resubmission either receives an
evaluation or, once the retry budget is spent, ends the parent in
terminal $\REJECT$. Two consecutive non-progressing resubmissions,
including duplicates, trigger $\HUMANGATE$. Of any two consecutive
resubmissions, at least one therefore consumes one of the $K+1$
evaluations or ends the parent. The property
guarantees finite \emph{termination}, not eventual commit: an intent
infeasible in the world should terminate as a rejection or an
escalation.

\subsection{Joint Contract Property}
\label{ssec:joint_contract}

The Joint Contract Property below rests on four enforcement
assumptions. \emph{K1 (complete mediation)}: every PRGA dispatch passes
through the executor, and every other write to a bound resource
advances its generation. \emph{K2 (bounded evaluation)}: every executor
evaluation runs under an enforced timeout. \emph{K3 (version-check
enforcement)}: the adapter performs a compare-and-set on the resource
lease and the expected policy/configuration generation. \emph{K4
(record emission)}: every terminal response includes a minimal
decision record.

\smallskip\noindent\textbf{Joint Contract Property.}\enspace
\emph{Under K1--K4, the PRGA contract jointly provides invariant
non-bypass of the specified hard-invalid cases; at most $K+1$ executor
evaluations per parent transaction, each under an enforced timeout; and
a minimal decision record in every terminal response, with any full
record added after the decision.}

\smallskip\noindent
The joint property follows from the preceding enforcement rules. K1 and
K3 ensure that every PRGA dispatch passes through its executor's
admission checks and the INV-5 check. K2, together with the four
mechanisms of \Cref{ssec:bounded_resolution}, limits each parent
transaction to $K+1$ evaluations. K4 ensures that every terminal
response includes a minimal record, and by INV-4 any full record is
generated after the online decision.

\smallskip\noindent\textbf{Example.}\enspace
A cell-sleep intent gated for a local conflict commits only on approved
coordination evidence (INV-3); the commit is bound to the evaluated
generation, which another planner's slice-protection update then
advances; INV-5 refuses the release with a typed receipt. The resubmission makes
progress, as the bound resource generation has changed, and earns a new
evaluation charged to the parent's retry budget.

%% file: sections/6_calibration.tex
\section{Calibration and Implementation}
\label{sec:calibration}

\subsection{Data Partitioning and Parameter Settings}
\label{ssec:splits}

\leavevmode Evaluation data are kept separate from the data used to
fit, calibrate or select the quantities consumed online. The final test
below supports the retrieval-cost comparison of
\Cref{ssec:cost_sensitivity}, and
\Cref{ssec:rq1b_native,ssec:rq5_quota} evaluate on fresh seeds of their
own. Seeds are
partitioned by run into five disjoint roles, in pipeline order:
\emph{fitting}, \emph{model selection}, \emph{probability calibration},
\emph{envelope} (the per-bin one-sided limits), and
\emph{operating-point validation} (the final-test arms' knobs). Splitting is by run/seed rather than by epoch, so no
epochs from the same run appear in different data roles. A sixth block of
$10$ fresh seeds is the \emph{final test}, evaluated once; it carries all
$52$ use-case $\times$ scenario blocks. The threshold tables of
\Cref{ssec:rq5_quota}, the operating points of \Cref{ssec:rq1b_native}
and of the final test, and the protocols of both RFsim campaigns were
fixed before their evaluation data were opened.

\input{figures/TNSE_REV_TABLE_PARAMETER_RULES.tex}

Table~\ref{tab:parameter_provenance} states how each parameter was
determined and which arms use it; \Cref{ssec:risk} and
Table~\ref{tab:thresholds}(c) give the threshold contract's default values.

\subsection{Two-Model Calibration and Pre-Specified Selection Rule}
\label{ssec:calibration_pipeline}

The executor scores an intent twice through the same pipeline, once on
$\Czero$-visible state and once after $\Cone$ evidence arrives. A
monotone logistic model with non-negative coefficients on $[0,1]$-scaled
features is fitted on the fitting role, so ``staler, more conflicted,
more congested'' can never score as safer; a probability-calibration map
is fitted on the calibration role; and the per-bin one-sided
Clopper--Pearson limits, monotonized by a cumulative-maximum majorant,
are computed on the envelope role after the three binning rules and the
level $\alpha_{\mathrm{UCB}}$ were frozen.

The choice among candidate model variants and the two calibration maps
follows a pre-specified rule:
minimize mean binary log-loss on the selection seeds, then, within one
standard error of the minimum, prefer the most parsimonious candidate.
Applied per stage, it returned the main-effects model with the Platt
calibrator at both stages.

Every envelope coverage statement in this paper is per bin and action
class, pointwise and \emph{marginal}, with no multiplicity correction.
On the held-out operating-point-validation role, the pre-specified
per-bin check found the empirical unsafe rate above the envelope value
in one bin per stage, the same action class and bin at both stages:
$0.16$ against $0.1316$ on $100$ observations, one-sided binomial
$p = 0.238$.

\subsection{Baseline Implementations and Fidelity}
\label{ssec:baseline_fidelity}

The harness is a CPU-only Python implementation: a scenario engine
generates diurnal load patterns and decision epochs in the context of
3GPP TR~38.864~\cite{3gpp38864} and TS~28.541~\cite{3gpp28541},
executors return $\COMMIT$/$\GATE$/$\REJECT$ on $\Czero$ payloads,
two to four simulated verifiers per request vote
per~\eqref{eq:quorum}, and injectors add conflicting intents on
requested resources to every stream, and verifier faults to some scenario
blocks.
Each verifier vetoes with probability
$0.05+0.3\,\riskt+0.2\,\lambda_t$ at the engine's current load
$\lambda_t$; if it does not veto, it escalates with probability $0.05$
and otherwise approves. The quota experiment of \Cref{ssec:rq5_quota}
and the predictive-admission rerun of \Cref{ssec:rq1b_native} use
estimate-based admission; the other offline runs of
Algorithms~\ref{alg:c0_triage}--\ref{alg:gate_resolve} use simulation
look-ahead.

Eager-C1 runs \textsc{PRGA-Fixed}'s executor path with the risk
envelope disabled and fetches $\Cone$ for every intent; it is used as a
retrieval-cost reference (\Cref{ssec:cost_sensitivity}).

We implemented the six reference executors in
Table~\ref{tab:native_confirmation} from the cited designs. None uses
PRGA's thresholds. Except for Simplex-RTA, they never actuate an intent
with an undischarged blocking precondition. Simplex-RTA,
Shielded-Executor, SHIELD-RIC-style and Conflict-Only fetch no $\Cone$,
replace a refused action by its hold/restore target and hand refused
SLA escalations to \HUMANGATE. Simplex-RTA~\cite{simplex2023} refuses
when the worst targeted load predicted from the action's declared effect
exceeds $1-h-\phi(a)$, where $h$ is the recoverable-set headroom.
Shielded-Executor~\cite{alshiekh2018shielding} refuses when the current
worst targeted load exceeds $1-m-\phi(a)$, where $m$ is the
load-uncertainty band. In both rules, the margin $\phi(a)$ is our
convention.
SHIELD-RIC-style refuses when a port of the source's
filter~\cite{dong2026shieldric}, ending in the violation gate
$\epsilon_s$ on a feature-count violation estimate, denies the
decision's one candidate. The filter's uncertainty cap and catalog ban
never trigger here. Conflict-Only~\cite{adamczyk2023oranconflict}
refuses when an active intent holds a targeted resource.
CAIF-style~\cite{bimo2026caif}, a schematic mapping, fetches $\Cone$,
pays for an in-path $\Ctwo$ audit write that our mapping adds,
rejects on a failed contract clause or a veto, and hands off to
$\HUMANGATE$ on missing information, no votes, an escalating vote or
too few approvals. Commit-Rollback~\cite{clemm2022rfc9315} actuates,
then rolls back on a majority refutation or an observed conflict;
an unsafe actuation counts as risk even when rolled back.

\subsection{Fairness Protocol}
\label{ssec:fairness}

\smallskip\noindent\textbf{Common random numbers.}\quad
Each (use case, scenario block, seed) triple generates one scenario
stream that every arm consumes, with pairing defined at the run/seed
level on trajectory metrics.

\smallskip\noindent\textbf{Metrics and risk scales.}\quad
Safe-Commit Yield and the unsafe-commit rate are the fractions of
candidate intents that the executor commits and whose evaluation-truth
label (\Cref{ssec:information_structure}) is safe and unsafe,
respectively. $\varepsilon_a$ is a \emph{design-time}, envelope-scale, per-class risk
budget of the contract, $\varepsilon_{\mathrm{eval}}$ an
\emph{evaluation-time}, outcome-scale, run-level cap.

\smallskip\noindent\textbf{Unsafe-epoch rate and UC1 action score.}\quad
In \Cref{ssec:rq5_quota,ssec:rq1b_native}, a seed's unsafe-epoch rate
is the fraction of epochs with an unsafe executed action in each run,
averaged equally over the run configurations in the evaluated
population. The risk test compares these rates
with a common cap of $5\%$ of epochs by a Holm-corrected Hoeffding--KL
test, which treats master seeds as independent units. The UC1 KPI is
the \emph{action-weighted UC1 energy score} (UC1 action score): each
executed action books a synthetic percentage drawn by action type,
largest for a cell sleep, and an epoch without execution books zero; it
averages over all epochs, is not integrated energy, and counts repeated
sleeps and wakes. A sleep lowers only its own cell's utilization;
neighbor offloading is not modeled.

\smallskip\noindent\textbf{Operating-point selection.}\quad
For the final-test runs, each arm's operating point is selected
independently on the validation role to maximize Safe-Commit Yield
subject to an upper confidence bound on the
unsafe-commit rate at or below $\varepsilon_{\mathrm{eval}} = 0.05$ and
the same QoS constraints.
The rule of \Cref{ssec:rq1b_native} selects
Table~\ref{tab:native_confirmation}'s points, and~\eqref{eq:q} the table
entries of \Cref{ssec:rq5_quota}. In each comparison one risk cap
bounds every arm's risk without equalizing it or constraining which
invariants an arm enforces.

\subsection{Emulation Path and Downstream Actuation}
\label{ssec:rfsim_impl}

Release protection and recovery after refusal are evaluated on an
AgentRAN-integrated OAI RFsim stack. It is a software-radio emulation
of the RAN driven by scripted request sources through a gateway executor separate from the
offline engine or, in the resend study of \Cref{ssec:rq4_rfsim}, by a
scripted controller. Downstream actuation goes through a \emph{testbed
REST adapter}, at which the \mbox{INV-5} fence is \emph{adapter-level atomic
conditional actuation}. PRB-mask writes are read back independently;
other writes report acknowledgment only. Wall-clock numbers are from
Jetson Orin (Arm), the resend study's from x86.

On the RFsim adapter path, an in-window race against a stale expected
generation was tested in $14$ armed attempts, of which $13$ were valid.
All $13$ valid trials were refused atomically, with no invalid
downstream actuation. All $13$ clean controls were applied. The
final-test runs and those of Table~\ref{tab:closed_loop} bind no
resource generation and therefore do not exercise the fence. The
fence's pooled p99 execution time provides the reserve
$L_{\mathrm{fence}}$ for the quota study in \Cref{ssec:rq5_quota}.

\emph{Decision records.} Every terminal response includes a typed
minimal decision record. The gateway executor also persists this record
locally before returning the response, with the append cost included in
attempt latency. Any full $\Ctwo$ record is written after the
decision.

\subsection{O-RAN Placement}
\label{ssec:a1_pattern}

\nobreak
PRGA sits at SMO / Non-RT~RIC supervisory timescales inside the
supervisory workflow, replacing no O-RAN
interface~\cite{polese2023oran,oran2024wg2}. Its terminal decisions are
\emph{pre-dispatch}. A $\COMMIT$ authorizes an A1-P Policy Create/Update
or, for the actions realized over O1 in
Table~\ref{tab:action_catalog}(b), an O1 configuration change. A
$\REJECT$ is a local executor/rApp response that emits no downstream
actuation request. For A1, policy feedback applies only to
already-created policy instances. A Near-RT~RIC create failure (HTTP~400
validation, 409 conflict) is a distinct downstream rejection point, and
post-create enforcement changes map to A1 Notify Policy
Status~\cite{etsi2025a1gap,etsi2025a1ap}. $\Cone$ evidence comes from
SMO/R1/O1 services or the executor-local registry, with A1 Enrichment
Information used only where its semantics match; verifier results are
external and $\Ctwo$ storage an SMO audit service, neither native A1.

%% file: figures/TNSE_REV_TABLE_PARAMETER_RULES.tex
\begin{table*}[t]
\centering
\caption{Parameter settings and how they were determined.}
\label{tab:parameter_provenance}
\footnotesize
\setlength{\tabcolsep}{3.0pt}
\begin{tabular}{@{}>{\raggedright\arraybackslash}p{\dimexpr0.15\textwidth-\tabcolsep\relax}>{\raggedright\arraybackslash}p{\dimexpr0.55\textwidth-2\tabcolsep\relax}>{\raggedright\arraybackslash}p{\dimexpr0.30\textwidth-\tabcolsep\relax}@{}}
\toprule
\textbf{Parameter} & \textbf{Setting and source} & \textbf{Used by} \\
\midrule
Score weights and contract thresholds & Hand-set defaults: weights in \eqref{eq:risk_score} and thresholds in Table~\ref{tab:thresholds}(c). & \textsc{PRGA-Fixed} and the closed loop of Table~\ref{tab:closed_loop}. Quota arms use only $\deltastale$, $m_d$ and $\beta_{\min}$. \\
Risk model and probability map & Monotone logistic model with non-negative coefficients, fitted on fitting seeds; Platt map fitted on calibration seeds. The one-standard-error rule on selection seeds chooses the main-effects model with Platt calibration at both stages. & \textsc{PRGA-Fixed} and quota arms use separate fits with the same rule and selected form (\Cref{ssec:calibration_pipeline}). \\
Envelope $\bar U_{a,b}$ & One-sided Clopper--Pearson upper limits by action class and score bin, with $\alpha_{\mathrm{UCB}}=0.05$, computed on envelope seeds and monotonized by \eqref{eq:majorant}. & One envelope per fit. In the quota fit, it passes the low and medium classes and refuses the high class. \\
Risk budgets $\varepsilon_a$, $\varepsilon_a^{\mathrm{deg}}$ & Base $\varepsilon_a=0.20/0.10/0.05$ for low/medium/high classes, placed against the envelopes' score support so the multiplier grid spans always- to never-commit, fixed before operating-point selection and the final test; $\varepsilon_a^{\mathrm{deg}}=\varepsilon_a/2$. & Multipliers selected per arm (\Cref{ssec:fairness,ssec:rq1b_native}); quota arms use base values. $\varepsilon_a^{\mathrm{deg}}$ applies only to degraded commits. \\
Local probability and divergence tolerance & Local unsafe probability: fixed-form main-effects logistic model with a Platt map. Tolerance: the $0.95$ quantile of $|$declared $-$ local$|$ on honest declarations from validation seeds, replacing $\epstrust$. & Quota arms; exceeding the tolerance makes $\Cone$ evidence mandatory. \\
Certificate latency leg & Transfer time of $\widehat{B}_{\Cone}$ at the link rate, plus the $0.95$ verifier-delay quantile by use case and action (fitting seeds), plus the fence reserve of $1646$\,ms, the pooled p99 fence execution time at the adapter over passes and refusals (\Cref{ssec:rfsim_impl}). & Quota arms; the predictive-admission rerun of \Cref{ssec:rq1b_native} omits the fence reserve. \\
Quota and byte settlement & $\widehat{B}_{\Cone}=1249$\,B (schema bound); $\kappa\in\{0.25,0.5\}$ of the reference budget. Admission requires the full reservation; delivered bytes are charged and unused reserved bytes refunded. The remaining balance expires at period end. & Quota arms. \\
Threshold candidates $\Theta$ and selection (Q) & $20$ candidates per use case: $+\infty$ and evenly spaced base-fetch score quantiles from quota-free fitting runs at four of the declared start phases. Each entry is enumerated on $14$ fitting seeds with $\bar R=1.30\%$; no candidate exceeded $\bar R_j$ on these fitting runs. & \textsc{GS8} and the level-class and phase tables used by \textsc{PRGA-PE-BC} and \textsc{PRGA-PE-BCI}. \\
Phase grid and profile & Declared phases $\{0,3,\ldots,21\}$\,h; $240$\,s estimation grid. Mean-score profiles by use case, action class and load-cycle hour are fitted on fitting seeds and interpolated between hour centers. & \textsc{PRGA-PE-BC} and \textsc{PRGA-PE-BCI}; online estimation as specified in \Cref{ssec:prga_ps}. \\
Switching points and policy form & $n_1=50$, $n_2=200$, set by a misclassification calculation before development. Policy form chosen among three declared forms on development seeds, disjoint from the fitting and evaluation seeds. & \textsc{PRGA-PE-BC} and \textsc{PRGA-PE-BCI} (the three segments of \Cref{ssec:prga_ps}). \textsc{PRGA-PE-BC} maps the estimate to the nearest declared phase. After $n_2$, \textsc{PRGA-PE-BCI} interpolates thresholds at the two declared phases bracketing the grid estimate. \\
\bottomrule
\end{tabular}
\end{table*}

%% file: sections/6_evaluation.tex
\section{Performance Evaluation}\label{sec:evaluation}

\subsection{Evidence Selection under a Shared Quota}
\label{ssec:rq5_quota}

\input{figures/TNSE_REV_TABLE_PHASE_CONFIRMATION.tex}

\begin{figure}[t]
\centering
\includegraphics[width=\columnwidth]{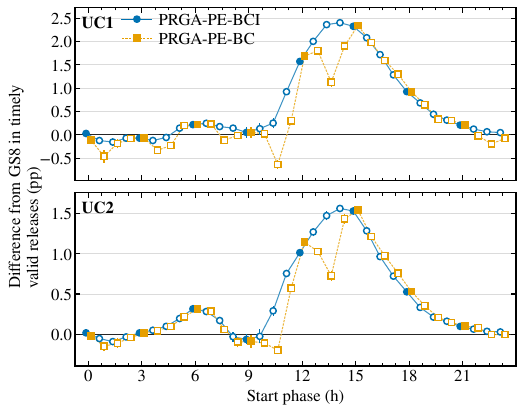}
\caption{Differences from \textsc{GS8} in timely valid release
rates across start phases for UC1 and UC2. Filled markers identify the
eight declared start phases.}
\label{fig:phase}
\end{figure}

Under the same quotas, risk tests and release checks,
\textsc{PRGA-PE-BCI} achieves the highest mean timely valid release rate
among the four arms in both use cases
(Table~\ref{tab:quota_confirmation}). The arms differ only in the rule
that selects requests for evidence acquisition. A run is one budget period of $2000$ epochs (about $11$\,h
simulated) under a static quota of $\kappa \in \{0.25, 0.5\}$ times the
pre-specified $192$\,B/epoch reference budget, and its diurnal load starts at one of
$32$ start phases $45$\,min apart, eight of them declared. The design and
the primary contrast, \textsc{PRGA-PE-BCI}~$-$~\textsc{PRGA-PE-BC}, with
improvement defined as a $95\%$ interval entirely above zero, were fixed
before any run on $300$ fresh master seeds; the unsafe-epoch rate, its
risk test and UC1's action score are defined in \Cref{ssec:fairness}.
All arms share the quota executor of \Cref{ssec:prga_ps} ($1249$\,B reservations,
settlement, release checks), the fixed fitted risk,
divergence and verifier-delay inputs (Table~\ref{tab:parameter_provenance})
and the benign reference stream, a diurnal load with low conflict
intensity, loose deadlines, and no injected stale state or verifier faults.
Release and unsafe-epoch rates are
measured directly on fresh seeds; the calibration assumptions are
discussed in \Cref{ssec:calibration_pipeline}.

\textsc{PRGA-PE-BCI} raises timely valid releases over arrival-order fetching by
$1.358$ and $0.916\pp$ in UC1 and UC2 ($5.52$ and $4.49\%$ of
\textsc{FIFO}'s rate) and over the offline-selected global threshold by $0.633$
and $0.412\pp$ ($2.50$ and $1.97\%$ of \textsc{GS8}'s rate). Over
\textsc{PRGA-PE-BC}, the primary contrast, it adds $0.176$
$[0.167, 0.185]$ and $0.067$ $[0.062, 0.072]\pp$; both intervals lie
above zero, also at each quota level. Every arm passes the risk test.
On average, each arm settles between $139.4$ and $142.9$\,kB of $\Cone$
per run. \textsc{PRGA-PE-BCI} settles less than \textsc{FIFO} and no
more than \textsc{GS8}, so its gains over them do not come from spending
more of the quota. Over \textsc{PRGA-PE-BC}, it settles slightly more of
the same quota (Table~\ref{tab:quota_confirmation}).

\smallskip\noindent\textbf{Where the gains arise.}\quad
Fig.~\ref{fig:phase} shows the differences of \textsc{PRGA-PE-BCI} and
\textsc{PRGA-PE-BC} from \textsc{GS8} at $32$ start phases, averaged
over $300$ fresh master seeds and both quota levels. The pointwise
$95\%$ $t$ intervals are mostly shorter than the markers. At a few start phases
\textsc{PRGA-PE-BCI} falls below \textsc{GS8}. The gap between its curve
and \textsc{PRGA-PE-BC}'s is the gain from interpolation, which on
average arises between the declared start phases.

The higher rate of timely valid releases is accompanied by changes in
the mix of executed actions and trade-offs among the network KPIs.
Against \textsc{GS8} and \textsc{FIFO}, \textsc{PRGA-PE-BCI} on average
reduces UC1 SLA-violation minutes and throughput degradation, while
lowering the UC1 action score by at most $0.0202$ points. In UC2, it
increases throughput, while slice-SLA satisfaction and Jain fairness
fall by at most $0.155\pp$ and $0.00032$, respectively. P1 counts
releases, not their modeled effect (\Cref{ssec:objective_scope}), and
the threshold withholds fetches by planner score, so a policy that
releases more intents can release a different mix of them.

\subsection{Release Protection and Recovery}
\label{ssec:rq4_rfsim}

On the emulation path, the reservation guard refused every observed
attempt to send during another source's exclusive lease, while
reducing the completion rate of non-targeted requests. After a
version-mismatch refusal, an immediate resend delivered more transfers
on time than waiting and required more base-station (gNB) commands
(Table~\ref{tab:e3_formal}).

\input{figures/TNSE_REV_E3_FORMAL_RESULT.tex}

\smallskip\noindent\textbf{Receiver-side recovery after refusal.}\quad
In a second paired campaign, one immediate resend after a refusal
raised on-time receiver delivery from $77.3\%$ to $89.9\%$ relative to
waiting for the next decision. The paired mean gain was $1.133$
transfers per block of $9$ ($95\%$ $t$ interval $[0.678, 1.589]$;
$45$ protocol-valid blocks; part~(b) of Table~\ref{tab:e3_formal}).
This gain came at $3.133$ more gNB PRB-mask commands per block ($[1.137, 5.130]$), against $55.3$ when waiting.

Both arms used the same scripted controller for the PRB mask of a cell serving one user
equipment (UE): at each arrival and every $1$\,s it chose, from dead-reckoned progress
($\Czero$), the smallest of three PRB levels its estimate said met the deadlines, and wrote
only changed masks through the fence; a tenant's unfenced PRB borrowing
advanced the mask generation, so the fence refused stale writes.
A transfer was on time if the receiver held all of it within $0.3$\,s plus $1.2$ times
its size over the fully open delivery rate after its arrival.
After a refusal both arms read the state back; only the resend arm recomputed under the
same rule and resent at most once, reading no typed receipt.
Every refusal was a generation mismatch; all $343$ resends committed.
Neither arm applied a write with a superseded generation or outside
the allowed PRB range, or sent one after its intent deadline.

\subsection{Bounded Resolution}
\label{ssec:rq4prime_closed_loop}

\input{figures/TNSE_REV_TABLE_CLOSED_LOOP.tex}

At the default retry budget $K=3$, bounded resolution terminates every
evaluated parent transaction and safely commits all parents in the
repairable and temporarily infeasible workloads
(Table~\ref{tab:closed_loop}).
Offline, scripted replanners resubmit rejected UC1 actions under the UC1
thresholds in three workload families: repairable, temporarily
infeasible and permanently infeasible parents. Naive replay, an
unbounded reference, resubmits a rejected intent unchanged;
reason-aware replanning reads the typed receipt, repairs from it and
obeys its terminal flag, but has no retry budget, deduplication,
progress gate or backoff, which Bounded PRGA adds
(\Cref{ssec:bounded_resolution}). Bounded PRGA needs at most $3$ and $2$
evaluations for the repairable and temporarily infeasible parents
against the bound $K+1$ and ends half of the permanently infeasible
parents by human escalation: those vetoed with an empty admissible
interval, whose resubmissions make no progress. Reason-aware replanning also commits
every repairable and temporarily infeasible parent but keeps repairing
those vetoed parents until the harness cap of $24$ submissions stops it,
and naive replay leaves
every repairable and permanently infeasible parent unresolved. Backoff
makes Bounded PRGA slower than reason-aware replanning on repairable
parents.

\subsection{Independent Modules under One Risk Cap}
\label{ssec:rq1b_native}

\input{figures/TNSE_REV_TABLE_NATIVE_CONFIRMATION.tex}

Table~\ref{tab:native_confirmation} evaluates the service, risk, and
online cost of \textsc{PRGA-Fixed} and six reference executors under a
common risk cap. These runs impose no evidence quota and do not exercise
version fencing;
\Cref{ssec:rfsim_impl,ssec:rq5_quota,ssec:rq4_rfsim,ssec:rq4prime_closed_loop}
evaluate the functions that distinguish PRGA.

In Table~\ref{tab:native_confirmation}, \textsc{PRGA-Fixed} and six
modules (\Cref{ssec:baseline_fidelity}) were each tuned over their respective
parameter grids on $300$ operating-point selection seeds: among the points whose
unsafe-epoch rate passed a Holm-corrected Hoeffding--KL test at a
common cap of $5\%$ of epochs ($42$ of $68$ points qualified), the one with the highest mean
primary KPI was selected (ties: fewer online bytes, then a fixed order),
frozen, and evaluated once on $300$ fresh seeds, where all $11$ selected
points passed the risk test again.

In UC1 \textsc{PRGA-Fixed} is higher than Simplex-RTA, CAIF-style and
Conflict-Only by $2.01$, $0.69$ and $1.51$ score points at a higher
observed risk, with the most SLA-violation minutes of the five.
Shielded-Executor is better than \textsc{PRGA-Fixed} at this pair of
points on the modeled service and cost metrics at a close observed risk:
$0.71$ more score points (higher on each of the $300$ seeds), $207$ fewer online bytes per
epoch, no handoffs. In UC2 \textsc{PRGA-Fixed} is $0.20$ and
$0.06\pp$ above SHIELD-RIC-style and CAIF-style at a higher observed
risk, $0.03\pp$ below Conflict-Only and $0.25\pp$ below the two
highest-SLA points, which barely act and hand off $15.1\%$ of epochs.
Each primary-KPI improvement of \textsc{PRGA-Fixed} reported above
passes a one-sided paired $t$ test over the fresh seeds,
Holm-corrected over all $12$ module and use-case pairs.

Rerun post hoc at its frozen points with predictive
instead of look-ahead fetch admission, \textsc{PRGA-Fixed} shows no
detected primary-KPI difference in either use case (UC2: $95\%$ interval
$[-0.0042, +0.0014]\pp$), and each of its primary-KPI differences from
the modules keeps its sign and its test outcome; the rerun also passes
the risk test in both use cases.

\subsection{Retrieval Cost and Sensitivity of the Threshold Contract}
\label{ssec:cost_sensitivity}

Selective retrieval lowers \textsc{PRGA-Fixed}'s retrieval bytes and
median attempt latency against Eager-C1, which runs
\textsc{PRGA-Fixed}'s executor path with the risk envelope disabled and
fetches $\Cone$ for every intent (\Cref{ssec:baseline_fidelity}). In
every final-test run, Eager-C1's decision counts and recorded QoS aggregates
match those of \textsc{PRGA-Fixed}. On the final test (\Cref{ssec:splits}), \textsc{PRGA-Fixed} fetches $\Cone$ on
$0.8065$ of attempts; its online retrieval bytes per attempt have a
block median of $600.3$ against $652.8$\,B, a median blockwise reduction
of $7.43\%$. Its median attempt latency, from arrival to the terminal
executor decision, is $1532$ against $1809$\,ms because it fetches less
often, not faster: $3.02$\,ms without a fetch and $1789$\,ms with one.
With no verification server or queue shared between requests, a skipped
fetch frees capacity for later requests only under a quota
(\Cref{ssec:rq5_quota}).

The sensitivity analysis examines \textsc{PRGA-Fixed} at the decision
and run levels. At the decision level, recorded validation requests are
re-routed under score weights sampled uniformly on the simplex and
varied commit and reject thresholds. The resulting unsafe-commit rates
are $0.0072$ at the median and $0.0244$ at the 95th percentile.

At the run level, \textsc{PRGA-Fixed}'s yield varies more in UC2 than in UC1
across the tested threshold and staleness settings.
We reran \textsc{PRGA-Fixed} at the base risk
budgets on validation seeds of the benign reference stream over $100$
Latin-hypercube settings per use case. The settings scale the commit and
degraded-commit thresholds by \mbox{$\times0.5$--$1.5$}, the reject threshold
by \mbox{$\times0.8$--$1.1$} and $\deltastale$ by \mbox{$\times0.5$--$2$} of their
Table~\ref{tab:thresholds}(c) defaults. \textsc{PRGA-Fixed}'s seed-mean
yield has 5th--95th percentiles of $0.301$--$0.319$ (UC1) and
$0.203$--$0.326$ (UC2).
Across these settings, the per-setting upper confidence bound on the
unsafe-commit rate is at most $0.0143$ (UC1) and $0.0110$ (UC2), below
$\varepsilon_{\mathrm{eval}} = 0.05$.

%% file: figures/TNSE_REV_TABLE_PHASE_CONFIRMATION.tex
\begin{table}[t]
\centering
\caption{Evidence allocation under a shared $\Cone$ quota.}
\label{tab:quota_confirmation}
\footnotesize
\setlength{\tabcolsep}{4pt}
\newcommand*{\ucrow}[1]{\multicolumn{5}{@{}l}{\emph{#1}}}
\begin{tabular}{@{}lrlrr@{}}
\toprule
Arm & Releases & \textsc{PRGA-PE-BCI} $-$ arm & Unsafe & $\Cone$ \\
\midrule
\ucrow{UC1} \\
FIFO & $24.59$ & $+1.358$ $[1.335, 1.381]$ & $0.870$ & $142.9$ \\
GS8 & $25.31$ & $+0.633$ $[0.616, 0.651]$ & $0.838$ & $141.9$ \\
\textsc{PRGA-PE-BC} & $25.77$ & $+0.176$ $[0.167, 0.185]^{\dagger}$ & $0.825$ & $139.4$ \\
\textsc{PRGA-PE-BCI} & $25.94$ &  & $0.830$ & $141.1$ \\
\midrule
\ucrow{UC2} \\
FIFO & $20.43$ & $+0.916$ $[0.899, 0.933]$ & $0.832$ & $142.9$ \\
GS8 & $20.93$ & $+0.412$ $[0.402, 0.422]$ & $0.813$ & $141.9$ \\
\textsc{PRGA-PE-BC} & $21.27$ & $+0.067$ $[0.062, 0.072]^{\dagger}$ & $0.804$ & $141.1$ \\
\textsc{PRGA-PE-BCI} & $21.34$ &  & $0.806$ & $141.9$ \\
\bottomrule
\end{tabular}
\\[2pt]\raggedright\scriptsize
 Releases: timely valid releases, \% of epochs; differences in percentage points (pp); Unsafe: unsafe-epoch rate, \% of
epochs; $\Cone$: settled $\Cone$, kB per run. Means over $300$ fresh master seeds, with start phases and the two
static quotas weighted equally; differences seed-paired, with two-sided $95\%$ $t$ intervals. $^\dagger$Primary
contrast, fixed before the runs with its criterion, an interval entirely above zero; other differences are
descriptive.
\end{table}

%% file: figures/TNSE_REV_E3_FORMAL_RESULT.tex
\smallskip\noindent\textbf{Reservation guard.}\quad
A paired campaign toggles only the reservation guard. Both arms ran the same INV-5 check at the
adapter. In this campaign, no lease or policy version was configured for the adapter's INV-5 check, so
it compared only the resource generation. Foreign leases did not advance that generation. With the guard,
the adapter atomically reserves a write's physical
write set and refuses the write while an exclusive lease or another write's
reservation overlaps that set; while the reservation is held, no new
overlapping exclusive lease is accepted. Each of the $72$ pairs uses a $240$\,s window per arm at
the pilot-selected conflict density $0.2$.
The primary event is physical sending during a foreign exclusive lease's
recorded valid lifetime.

\begin{table}[t]
\centering
\caption{Release protection and recovery in two paired RFsim experiments.}
\label{tab:e3_formal}
\footnotesize
\setlength{\tabcolsep}{3pt}
\begin{tabular}{@{}lrrr@{}}
\toprule
\emph{(a) Guard, $72$ pairs} & OFF & ON & OFF $-$ ON \\
\midrule
$E/U$ windows & $66/1$ & $0/0$ & \\
\multicolumn{3}{@{}l}{Non-targeted completion (pp)} & $+8.05$ $[5.74, 10.37]$ \\
\midrule
\emph{(b) Resend, $45$ blocks} & Wait & Resend & \shortstack[r]{Resend $-$ Wait\\per block} \\
\midrule
On time (of $405$) & $313$ ($77.3\%$) & $364$ ($89.9\%$) & $+1.133$ $[0.678, 1.589]$ \\
gNB commands & $55.3$ & $58.4$ & $+3.133$ $[1.137, 5.130]$ \\
Refusal to commit (s) & $0.526$ & $0.175$ & --- \\
\bottomrule
\end{tabular}
\\[2pt]\raggedright\scriptsize
(a) $E/U$: windows with an event/unknown outcome under the primary rule; completion of requests not
generated as conflicts, mean over the $71$ fully determined pairs.
(b) Protocol-valid blocks of $9$ tight-deadline transfers per arm; on-time
totals, gNB PRB-mask commands per block, median time from a refusal during a transfer to the arm's next commit.
Differences: means with two-sided $95\%$ $t$ intervals, approximate in (a).
\end{table}

Without the guard (part~(a) of Table~\ref{tab:e3_formal}), $66$ of the $72$ windows had a primary event and $1$ was unknown after a
collection failure. With it, no window had a primary event, and all $259$ observed opportunities, attempts to send while another
source's exclusive lease overlapped the write set, were refused. The guard lowers the completion of
non-targeted requests, those not generated as conflicts
(part~(a)); the requests that succeed without the guard include lease-invalid releases, and the
guard's reservation check, which does not distinguish lease ownership, also refused
$368$ requests under their own source's exclusive lease.

%% file: figures/TNSE_REV_TABLE_CLOSED_LOOP.tex
\begin{table}[t]
\centering
\caption{Resolution after rejection, offline: parent transactions by workload and replanning strategy.}
\label{tab:closed_loop}
\scriptsize
\setlength{\tabcolsep}{2.4pt}
\begin{tabular}{@{}llcccc@{}}
\toprule
\textbf{Workload} & \textbf{Strategy} & \textbf{Safe} & \textbf{Unresolved} & \textbf{Eval.} & \textbf{Median} \\
 & & \textbf{commit} & \textbf{at cap} & \textbf{max} & \textbf{time (ms)} \\
\midrule
Repairable & Bounded PRGA & 1.00 & 0.00 & 3 & 697.5 \\
 & Reason-aware replan. & 1.00 & 0.00 & 3 & 497.5 \\
 & Naive replay & 0.00 & 1.00 & 24 & n.r. \\
\addlinespace[1pt]
Temp. infeasible & Bounded PRGA & 1.00 & 0.00 & 2 & 1735.2 \\
 & Reason-aware replan. & 1.00 & 0.00 & 2 & 1735.2 \\
 & Naive replay & 1.00 & 0.00 & 16 & 1531.6 \\
\addlinespace[1pt]
Perm. infeasible & Bounded PRGA & 0.00 & 0.00 & 2 & 473.2 \\
 & Reason-aware replan. & 0.00 & 0.50 & 24 & n.r. \\
 & Naive replay & 0.00 & 1.00 & 24 & n.r. \\
\bottomrule
\end{tabular}
\\[2pt]\raggedright\scriptsize
 For each workload family, all three strategies are evaluated on the same $720$ parent transactions. \emph{Safe commit}: the share ending in a $\COMMIT$ labeled safe. \emph{Unresolved at cap}: the share stopped by the harness cap of $24$ submissions before $\COMMIT$, terminal $\REJECT$ or $\HUMANGATE$. Eval.\ max: most executor evaluations of any parent, bounded by $K+1$ ($K = 3$) for Bounded PRGA only. Median time: from a parent's first submission to its $\COMMIT$, terminal $\REJECT$ or $\HUMANGATE$; n.r.: at least half the parents unresolved. Retry-budget sweep, repairable family: safe-commit rate 0.9347 at $K = 1$, with at most 2 evaluations, and 1.0000 at $K = 3$ and $K = 5$.
\end{table}

%% file: figures/TNSE_REV_TABLE_NATIVE_CONFIRMATION.tex
\begin{table}[t]
\centering
\caption{Model comparison of \textsc{PRGA-Fixed} with independent reference executors under one
risk cap.}
\label{tab:native_confirmation}
\footnotesize
\setlength{\tabcolsep}{3.5pt}
\begin{tabular}{@{}lrrrrr@{}}
\toprule
\textbf{Selected point} & \textbf{Risk} & \textbf{KPI} & \textbf{Act.} & \textbf{Hand.} & \textbf{Bytes} \\
\midrule
\multicolumn{6}{@{}l}{\emph{UC1 energy saving; KPI = UC1 action score}} \\
\textsc{PRGA-Fixed} ($\times3$) & 0.97 & 3.15 & 32.5 & 5.0 & 599 \\
Simplex-RTA ($h{=}0.6$) & 0.23 & 1.13 & 14.4 & 0.0 & 392 \\
Shielded-Executor ($m{=}0.25$) & 0.97 & 3.86 & 44.3 & 0.0 & 392 \\
CAIF-style & 0.79 & 2.46 & 25.3 & 8.1 & 1245 \\
Conflict-Only & 0.61 & 1.63 & 16.4 & 0.0 & 392 \\
\multicolumn{6}{@{}l}{SHIELD-RIC-style, Commit-Rollback: no risk-qualified point} \\
\midrule
\multicolumn{6}{@{}l}{\emph{UC2 slice-SLA protection; KPI = slice-SLA satisfaction (\%)}} \\
\textsc{PRGA-Fixed} ($\times0.75$) & 0.96 & 85.34 & 27.5 & 4.5 & 602 \\
Simplex-RTA ($h{=}0.8$) & 0.00 & 85.59 & 0.0$^\ddagger$ & 15.1 & 410 \\
Shielded-Executor ($m{=}0.8$) & 0.00 & 85.59 & 0.0$^\ddagger$ & 15.1 & 410 \\
SHIELD-RIC-style ($\epsilon_s{=}0.05$) & 0.62 & 85.14 & 19.4 & 15.1 & 410 \\
CAIF-style & 0.86 & 85.29 & 22.1 & 7.0 & 1264 \\
Conflict-Only & 0.70 & 85.37 & 16.1 & 12.6 & 410 \\
\multicolumn{6}{@{}l}{Commit-Rollback: no risk-qualified point} \\
\bottomrule
\end{tabular}
\\[2pt]\raggedright\scriptsize
 Knobs in parentheses (\Cref{ssec:baseline_fidelity}; $\times$: \textsc{PRGA-Fixed}'s envelope
multiplier). Means over $300$ fresh master seeds. Risk (unsafe-epoch rate),
Act.\ (candidate actuation) and Hand.\ (handoff, ending the automated attempt with no modeled operator
effort, delay or outcome) in \% of all epochs; Bytes: online bytes per epoch. Differences in the text are
seed-paired means of unrounded values. $^\ddagger$Nonzero:
$68$ and $158$ candidate actuations in $15.6$ million epochs.
\end{table}

%% file: sections/7_discussion.tex
\section{Discussion and Limitations}\label{sec:discussion}

\nobreak
Acquiring evidence commits resources that later requests may need.
Under a shared quota, the value of a fetch therefore depends on both the
current intent and the requests that follow. The quota experiments show
that allocating evidence using the estimated load phase can increase
average timely valid releases while all policies retain the same
admission and release checks.

The recovery experiments show that refusing an invalid release and
completing the control task are separate concerns. In the RFsim resend
study, the same fence refused every stale write in both arms, while
immediate resend improved on-time delivery at additional command cost.
In the offline study, receipt-guided repair recovered the repairable
cases, and progress checks with a shared retry budget brought
permanently infeasible parents to a terminal decision.

PRGA targets SMO/Non-RT~RIC supervisory timescales. Its RFsim evaluation
uses scripted request sources or a scripted controller and a testbed
REST adapter. The remaining experimental results come from the offline
model. The online risk estimate can be miscalibrated
(\Cref{ssec:calibration_pipeline}). Future work includes evaluation on
operator traces, online phase inference for changed load shapes, and
concurrent LLM planners that repair refused intents from typed receipts.

%% file: sections/8_conclusion.tex
\section{Conclusion}\label{sec:conclusion}

\nobreak
PRGA couples selective evidence acquisition, release authorization, and
bounded resolution at the executor boundary between agent-generated
supervisory intents and network actuation. Its contract preserves
mandatory evidence requirements even when resources run short, rechecks
approvals against bound resource versions and leases before release,
and permits reevaluation only on receipt-defined progress within the
parent transaction's retry budget.
Under a common risk cap, comparisons with reference executors
implemented from independent designs characterize the trade-offs among
service performance, observed risk, and online cost. In modeled workloads with
stable daily load cycles, the proposed allocation policy increases
average timely valid release rates over arrival-order fetching and an
offline-selected global threshold under identical quotas and release
checks. OAI radio emulation demonstrates reservation-based release
protection at a lower completion rate for non-targeted requests. An
immediate scripted resend after a version-mismatch refusal delivers more
transfers on time than waiting, at modest additional command cost. These
results show that evidence allocation and responses to refusal can
improve timely supervisory execution without relaxing release checks.